\documentclass[final,pre,onecolumn,superscriptaddress,floatfix]{revtex4-1}
\usepackage{hyperref}
\usepackage{graphicx}
\usepackage{longtable}
\usepackage{psfrag}
\usepackage{color}
\usepackage{amsmath}
\usepackage{amssymb}
\usepackage{setspace}
\usepackage{amssymb,amsfonts,amsmath}
\usepackage[usenames,dvipsnames]{xcolor}
 \usepackage{multirow}
\usepackage{booktabs}
\newcommand{\uunder}[1]{{\bf{\color{red}#1}}}

\newcommand{\ooverrep}[1]{{\bf{\color{ForestGreen}#1}}}

\begin{document}

\title{Family-specific scaling laws in bacterial genomes. }

%%%\title{Family-specific evolutionary potentials in bacterial genomes. }

%%\title{The scaling laws in the bacterial genomic repertoires are
%%  family-specific. }

\author{Eleonora de Lazzari}
%
%%% 
\affiliation{Sorbonne Universit\'es, UPMC Universit\'e Paris 06, UMR 7238
Computational and Quantitative Biology, Genomic Physics Group, 15, rue
de l'\'{E}cole de M\'{e}decine, Paris, France }

\author{Jacopo Grilli}
%
%%% 
\affiliation{Department of Ecology and Evolution,
  University of Chicago, 1101 E 57th st., Chicago, IL, 60637, USA}

\author{Sergei Maslov}

\affiliation{Department of Bioengineering, Carl R. Woese Institute for Genomic Biology, University of Illinois at
Urbana-Champaign, Urbana, IL 61801, USA}

\author{Marco {Cosentino Lagomarsino}}
\affiliation{Sorbonne Universit\'es, UPMC Universit\'e Paris 06, UMR 7238
Computational and Quantitative Biology, Genomic Physics Group, 15, rue
de l'\'{E}cole de M\'{e}decine, Paris, France }
\affiliation{CNRS, UMR 7238, Paris, France} 
\affiliation{FIRC Institute of Molecular Oncology (IFOM), 20139 Milan, Italy}

\begin{abstract}
  Among several quantitative invariants found in evolutionary
  genomics, one of the most striking is the scaling of the overall
  abundance of proteins, or protein domains, sharing a specific
  functional annotation across genomes of given size.  The size of
  these functional categories change, on average, as power-laws in the
  total number of protein-coding genes.
  Here, we show that such regularities are not restricted to the
  overall behavior of high-level functional categories, but also exist
  systematically at the level of single evolutionary families of
  protein domains.
  Specifically, the number of proteins within each family follows
  family-specific scaling laws with genome size. Functionally similar
  sets of families tend to follow similar scaling laws, but this is
  not always the case. To understand this systematically, we provide a
  comprehensive classification of families based on their scaling
  properties. Additionally, we develop a quantitative score for the
  heterogeneity of the scaling of families belonging to a given
  category or predefined group.
  Under the common reasonable assumption that selection is driven
  solely or mainly by biological function, these findings point to
  fine-tuned and interdependent functional roles of specific protein
  domains, beyond our current functional annotations.  This analysis
  provides a deeper view on the links between evolutionary expansion of
  protein families and the functional constraints shaping the gene
  repertoire of bacterial genomes.
\end{abstract}

\maketitle

\section{Introduction}

As demonstrated by van Nimwegen~\cite{vanNimwegen03} and confirmed by
a series of follow-up
studies~\cite{Molina08,MOLINA09,cordero2009regulome,Grilli2012,charoensawan2010genomic},
striking quantitative laws exist for high-level functional categories
of genes. Specifically, the number of genes within individual
functional categories such as e.g. that of transcriptional
regulators~\cite{Stover2000,vanNimwegen03,Maslov2009} exhibit clear
power-laws, when plotted as a function of genome size measured in
terms of its number of protein-coding genes or, at a finer level of
resolution, of their constitutive domains.
In prokaryotes, such scaling laws appear well conserved across clades
and lifestyles~\cite{molina2009scaling}, supporting the simple
hypothesis that these scaling laws are universally shared by this
group.
%~ , despite the variety in lifestyles and habitats reflected
%~ by an equally large variation in detailed functional organization from
%~ genome to genome.

From the evolutionary genomics viewpoint~\cite{Koonin2011}, these laws
have been explained as a byproduct of specific ``evolutionary
potentials'', i.e., per-category-member rates of additions/deletions
fixed in the population over evolution. As predicted by quantitative
arguments, estimates of such rates correlate well with the category
scaling exponents~\cite{vanNimwegen03,Molina08}. A complementary point
of view~\cite{Maslov2009,Pang2011,Grilli2012} focuses on the existence
of universal ``recipes'' determining ratios of proteins between
different functions. Such recipes should mirror the ``dependency
structure'' or network operating within genomes as well as other
complex systems~\cite{Pang2013}. According to this point of view the
usefulness, and thus the occurence, of a given functional component
depends on the presence of a set of other components, which are
necessary for it to be operational.

Beyond functional categories, protein coding genes can be classified
in ``evolutionary families'' defined by the homology of their
sequences. Functional categories routinely contain genes from tens or
more of distinct evolutionary families. The statistics of gene
families also exhibits quantitative laws and regularities starting
from a universal distribution of their per-genome
abundance~\cite{Huynen98}, explained by evolutionary models accounting
for birth, death, and expansion of individual
families~\cite{Qian2001a,Karev2004,Cosentino09}.
%~ , describing family
%~ dynamics by moves such as gene loss, duplication and horizontal gene
%~ transfers.

% **************************************************************
% Keep this command to avoid text of first page running into the
% first page footnotes
\enlargethispage{-65.1pt}
% **************************************************************

While some earlier work connects per-genome abundance statistics of
families with functional scaling laws~\cite{Grilli2012}, the link
between functional category scaling and evolutionary expansion of gene
families that build them remains relatively unexplored. 
Clearly, selective pressure is driven by functional constraints, and
thus selection cannot in principle recognize families with identical
functional roles. On the other hand, slight differences in the
functional spectrum of different protein domains, and interdependency
of different functions can make the scenario more complex.
Thus, one central question is how the abundance of genes performing a
specific function emerges from the evolutionary dynamics at the family
level. 

Two alternative extreme scenarios can be put forward: (i) the
high-level scaling laws could emerge only at the level of functions, and
be ``combinatorially neutral'' at the level of the evolutionary
families building up a particular function, or, vice versa, (ii) they
could be the result of the sum family-specific scaling laws. In the
first scenario all or most of the families performing a particular
function would be mutually interchangeable. In the second scenario,
the evolutionary potentials would be family-specific and coincide with
family evolutionary expansion rates, possibly emerging from the
complex dependency structure cited above, and from fine-tuned
functional specificity of distinct families.
An intermediate possibility is that an interplay of constraints acts
on both functional and evolutionary families. The first test for the
feasibility of the second scenario is the existence of scaling laws
for individual families. 
Here, focusing on bacteria, and using protein domains to define
families, we present a clear evidence for family-specific scaling laws
with genome size. We show that the abundance of the families follows
power laws with genome size. Comparing functional categories with a
suitable null model, we show that family-specific exponents may
deviate significantly from the exponent of the associated functional
category.
We provide a comprehensive classification of families based on common
scaling exponents, which recovers the known functional associations as
well as revealing new ones, and may be used to detect possible misannotations. 
%such as the case of the functional category ``Protein modification''.	
%
Finally, we develop quantitative tools to measure the heterogeneity of
the scaling of families belonging to a given category or predefined
group of families.  
% This analysis shows a notable difference between the two families
% with ``most superlinear'' scaling, transcription factors and signal
% transduction.

\section{MATERIALS AND METHODS}

\subsection*{Data sources}
We considered bacterial proteomes retrieved from the SUPERFAMILY
(release 1.75 \textbf{downloaded in October 2014},
\cite{Gough01}) and PFAM (release 27.0
\textbf{downloaded in October 2014}, \cite{bateman2004pfam,Finn2014})
database. Evolutionary families were defined from the domain
assignments of 1535 superfamilies (SUPERFAMILY database) and 446 clans
(PFAM database) on all protein sequences in completed genomes. We
focused the analysis on the 1112 bacterial proteomes used as species
reference in the SUPERFAMILY database.
% (list in SI Table
%\textbf{.\textbackslash SupTables\textbackslash taxonomy}).
%%
For the functional annotations of the SUPERFAMILY data, we considered
annotation of SCOP domains as a scheme of 50 more detailed functional
categories, mapped to 7 more general function categories, developed by
C.~Vogel~\cite{vogel2006protein}.  PFAM clans were annotated on the
same scheme of 50 functional categories, using the mapping of clans
into superfamilies available from the PFAM website
\url{http://pfam.xfam.org/clan/browse#numbers}~\cite{finn2006pfam}.

\subsection*{Data analysis}
For each evolutionary domain family (or a functional category
consisting of multiple evolutionary families), genome sizes (measured
in the overall number of domains) were logarithmically binned. For
each bin we calculated mean and standard deviation of the given family
abundance (number of domains) within the bin.  The estimated scaling
exponent $\beta_{i}$ for family $i$ is the result of the non-linear
least squares fitting of the binned data weighted by the standard
error of family abundance. Genome size bins containing less than $10$
genomes were not taken into account.
To filter out the data that, due to low-abundance or rare families,
were affected by sampling problems, we considered three independent
parameters, (i) the ``occurrence'', i.e. the fraction of genomes where
family $i$ is present, $o_{i} = N_{G}^{(i)} / N_{G}$, where $N_{G}$ is
the total number of genomes in the sample, and $N_{G}^{(i)}$ is the
number of genomes where the family has non-zero abundance, (ii) the
goodness of fit index \begin{displaymath} s_{i} =
  \dfrac{1}{1+\sqrt{LS_{i}}}
\end{displaymath} where $LS_{i}$ is the error associated with the exponent
$\beta_{i}$, measured as the average squared deviation between the fit and the
logarithm of the empirical abundance (see SI sec.~S1), and (iii) the
Pearson correlation coefficient 
$\rho_{i}$ between the logarithm of the family abundance and the
logarithm of the 
genome size.
The index $s_i$ puts on the same ground families with different
exponents, but generally decreases as the scaling exponent increases,
in accordance with the growth of fluctuations in families with higher
exponents observed in ref.~\cite{Grilli2014}.  Hence, we decided to
use it only for low exponents, where the Pearson correlation is a bad
proxy of scaling.  We considered families with $s_{i} > 0.9$ and
$o_{i} > 0.6$ for exponents lower than 0.2, otherwise families with
$\rho_{i} > 0.4$ and $o_{i} > 0.6$ reducing the dataset to 357
superfamilies and 178 clans that satisfy both requirements. As shown
in Fig.~\ref{fig:corroccurence}A, $s_{i}$ and $o_{i}$ are not mutually
correlated across the genomes, implying that the two requirements are
in fact independent, the same is valid for $\rho_{i}$ and $o_{i}$, see
Fig.~\ref{fig:corroccurence}B.
We verified that the removed families with the procedure described
above do not influence the scaling of the category.  Supplementary
Fig.~\ref{fig:cat_cfr} reports the exponent of the category scaling
before the thresholding (where all the families are considered) and
after (where the domains belonging to the removed families are not
considered in the category scaling), showing that the values are
consistent for all the categories studied.

%%%%%
%%% Heterogeneity score
%%%
For each family within a given functional category, we defined a
``heterogeneity score'' $h_{i}$ as follows:
\begin{displaymath}
  h_{i} = |\beta_{c}-\beta_{i}|, 
\end{displaymath}
where $\beta_i$ and $\beta_c$ are,respectively, the scaling exponents
of family $i$ and functional category $c$.  The heterogeneity measure
for each functional category was defined as the average of the
per-family heterogeneity scores $h_i$:
\begin{displaymath}
  H_{c} = \dfrac{1}{F_{c}} \sum_{i} h_{i} ,
\end{displaymath}
where $F_c$ is the number of families in category $c$. 

The significance of the values found with this formula was assessed
against a null model assuming that the total abundance of a category
is distributed randomly across the associated families. The average
abundance (i.e. the fraction of domains belonging to a family averaged
over genomes) and occurrence (fraction of genomes where the family is
present) of each family are both conserved (note that these two
properties are uncorrelated in the data, hence we chose to conserve
both in the null model, assuming that they are independent, see
Supplementary Fig.~\ref{fig:occvsfreq}).

Given a genome $g$ with $n^{g}_c$ elements in the functional category
$c$, divided into $F^g_c$ associated families, we redistributed the $n^{g}_c$
members among the $F^g_c$ sets conserving the average relative abundance of
each family (see SI sec. S2).  A member of family $i$
belonging to category $c$ was therefore added with probability
\begin{displaymath}
p_{i,c} \propto
\begin{cases}
  \dfrac{1}{N_{g}^{(i)}} \sum\limits_{g'=1}^{N_{G}^{(i)}}
  \dfrac{n_{i}^{g'}}{\sum_{k \in c} n_{k}^{g'} } &, \ \  \mbox{if }n_{i}^{g}\neq
  0 \\
  0 &, \ \  \mbox{if }n_{i}^g=0 \ .
\end{cases}
\end{displaymath}
The resulting set of $F_{c}$ artificially built evolutionary
categories constrains the occurrence pattern and the average abundance
of the original ones. 
Scaling exponents for families in the null model are extracted with
the procedure described above.
Only functional categories containing domains from more than $10$
distinct families were compared to the null model.
All procedures were implemented as custom Python 2.7 scripts.

\section{RESULTS}

\subsection*{Families have individual scaling exponents, reflected by
  family-specific scaling laws}
  
We started by addressing the question of whether individual families
show scaling laws, and thus can be associated to specific scaling
exponents. In order to do so, we isolated domains belonging to the
same family across the sample of 1112 species-representative bacterial
proteomes and plotted their abundance against the total number of
domains in the corresponding proteome.

\begin{figure}[htp]
\centering
  \includegraphics[width=0.45\textwidth]{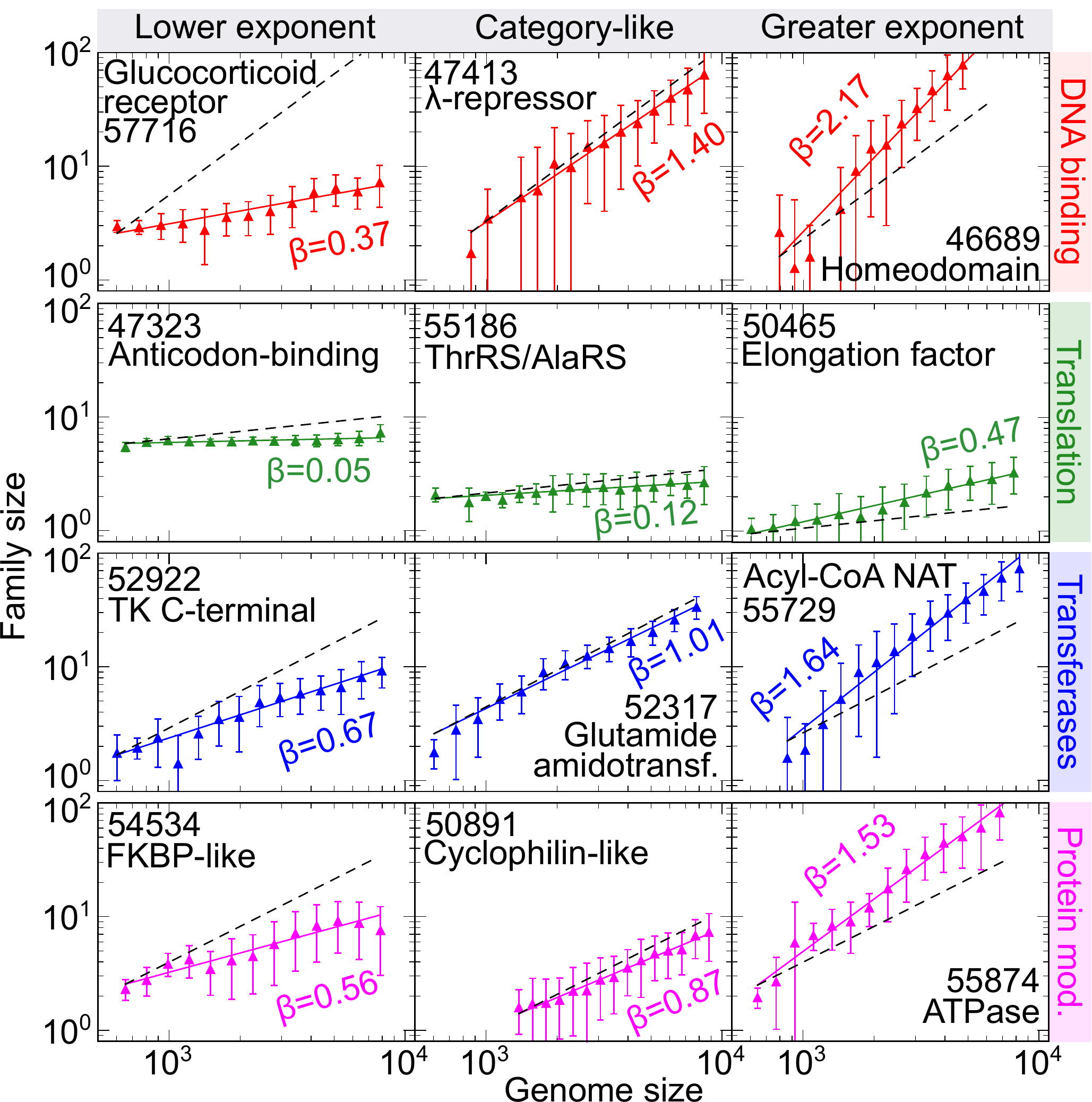}
  \caption{Families follow specific scaling laws, which may agree or
    deviate from the overall scaling of the functional category to
    which this family belongs.  The plots report the abundance of
    twelve different superfamilies as a function of the genome size
    (triangles are binned averages). The power-law fits (colored
    lines) are compared to the power-law fits of the functional
    category to which each family belong (dashed black lines). We
    display here examples from four functional categories: DNA binding
    (top row), Translation (second row from top), Transferases (third
    row from top) and Protein modification (bottom row). Families in
    the leftmost / rightmost column scale respectively slower/faster
    than their category means, families in the middle column have
    similar slope to the full category. Legends specify the SCOP
    superfamily id, family descriptive name and power-law exponent
    ($\beta_{i}$) from the fits. }
\label{fig:Fig1_fam_scaling}
\end{figure}

When the abundance is sufficiently high to overcome sampling problems,
most families show a clearly identifiable individual scaling when
plotted as a function of genome size. As an example,
Fig.~\ref{fig:Fig1_fam_scaling} shows the scaling of a set of chosen
families in four selected functional categories.  Additionally, some
low-abundance families that occur in all genomes with a very
consistent number of copies show definite scaling with exponents close
to zero~\cite{Grilli2014}, being clearly constant with size, with
little or no fluctuations.
 
Given that functional categories follow specific scaling laws, likely
related to function-specific evolutionary
trends~\cite{vanNimwegen03,Molina08}, there remain different open
possibilities for the behavior of the evolutionary families composing
the functional categories. One simple scenario is that family scalings
are family-specific, thus validating the existence of family
evolutionary expansion rates that are quantitatively different to the
one of their functional category. In the opposite extreme scenario the
scaling is only function-specific, and individual families performing
similar functions are interchangeable.  If this were the case, family
diversity in scaling exponent would be only due to sampling effects,
and the null model would fully reproduce the diversity in family
scaling observed in empirical data.
To address this question, we randomized the families within a category
conserving their occurrence patterns and the category average
abundance. The randomized families always show very similar scaling as
the one of the corresponding category (see Supplementary
Fig.~\ref{fig:nullmodel}). Hence, this analysis strongly supports the
existence of family-specific scaling exponents that do not simply
descend from the category scaling.

Fig.~\ref{fig:Fig1_fam_scaling} shows that the presence of
``outlier families'' is common among functional categories.
%
%In order to gain a systematic insight on the range of evolutionary
%potentials of the families composing each functional category, we
%compared the scaling exponents $\beta_{c}$ of the functional category
%with the exponents $\beta_{i}$ of the associated families.
%
In most categories, we found families where the deviations from the
category exponents is clear, beyond the uncertainty due to the errors
from the fits. Fig.~\ref{fig:Fig1_fam_scaling} shows some examples
where in each of the shown categories $\beta_{i}$ may be higher, lower
or comparable to $\beta_{c}$.
A table containing all the family and category exponents is
available as supplementary information (Supplementary Table~S1 and Table~S2).

Finally, we considered the correlation of family scaling exponents
with relevant biological and evolutionary parameters such as
foldability (quantified by size-corrected contact order,
SMCO~\cite{Debes2013}), the diversity of EC-numbers associated with
families (quantifying the functional plasticity of a given family),
selective pressure (quantified by the ratio of nonsynonimous to
synonymous $K_a/K_s$ substitution rates~\cite{Ndhlovu2015}) and
overall family abundance. The results are summarized in Supplementary
Table~\ref{tab:TableCorrelations}. Foldability and $K_a/K_s$ appear to
have little correlation with scaling exponents.  Instead, we found a
significant positive correlation of exponents with family abundance,
and both quantities are correlated with diversity of EC-numbers in
metabolic families. This suggests that, at least for metabolism,
functional properties of a fold play a role in family scaling, and
that beyond metabolism, abundance and scaling are, on average, not
unrelated.

\subsection*{The heterogenetiy in scaling exponents is
  function-specific.}

The analyses presented above support the hypothesis that functional
categories contain families with specific scaling exponents. Indeed,
the scaling exponents $\beta_i$ of the families can be significantly
different from the category exponent $\beta_c$ with deviations that
are much larger than predicted by randomizing the categories according
to the null model (see Supplementary Fig.~\ref{fig:nullmodel}).
 
In order to quantify this ``scaling heterogeneity'' of functional
categories, we computed for each family $i$ the distance between its
scaling exponent $\beta_{i}$ and the category exponent $\beta_{c}$
(see Methods). We defined an index $H_c$ quantifying the heterogeneity
of the scaling of the families within a category by averaging this
distance over the families associated to a given category $c$.

\begin{figure}[htp]
\centering
\includegraphics[width=0.48\textwidth]{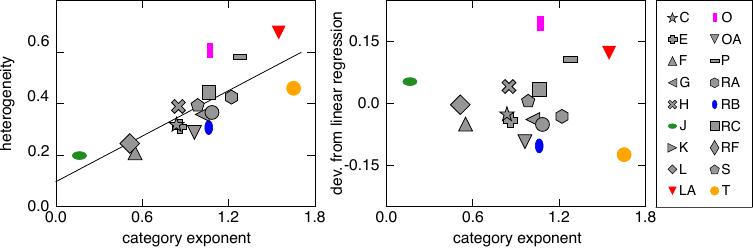}
\caption{ (A) Functional categories with faster scaling laws contain
  families with more heterogeneous scaling exponents. Heterogeneity is
  quantified by the mean deviation between the family scaling
  exponents and the category exponent. The plot reports heterogeneity
  scores for different functional categories, plotted as a function of
  the category exponents. The black line is the linear fit between
  heterogeneity and exponets (slope 0.3, intercept 0.1). (B)
  Comparison of heterogeneities subracted from the linear trend. By
  this comparison, the least heterogeneous categories are Signal
  Transduction (T) and Transferase (RB), and the most heterogenous are
  DNA Binding (LA) and protein modification (O). Translation (J) is
  slightly above the trend for its low exponent.  The legend (right
  panel) shows the association between symbols and category codes (see
  Supplementary Table~\ref{tab:TableS0sym} for the corresponding
  category name). }
\label{fig:Figure2}
\end{figure}

Figure~\ref{fig:Figure2}A shows the relation between the heterogeneity
$H_c$ and the category exponent $\beta_c$. Interestingly, these two
quantities are correlated, with categories with larger values of
$\beta_c$ being more heterogeneous.  This result can be intuitively
rationalized in terms of the degrees of freedom imposed by the
category exponent to the scaling of single families.  Categories with
small exponents are incompatible with extremely large fluctuations of
family exponents, while categories with larger exponents can contain
families with small $\beta_i$. Indeed, this trend of heterogeneity
with exponents is also observed in the null model, where the
hetereogenity of null categories is much smaller than empirical ones,
since all families tend to take the exponent category (Supplementary
Fig.~\ref{fig:nullmodel}).  
% (see Fig.~\ref{fig:Fig2_expVSrank}).
%This relation can be paralleled with the observation that the exponent
%of the family that deviates most from the null model in each category
%is linearly related with the category exponent (see Figure
%SXXX). \edl{a che figura è riferito??} \textbf{MCL CREDO A QUELLA DEL MASSIMO
%   FigSXrealExpmaxZexp} \edl{ok ma allora non possiamo dire 'null model'}

Figure~\ref{fig:Figure2}B allows a direct comparison of the
heterogeneity of different categories by subtracting the mean
trend. It is noteworthy that the Signal Transduction functional
category, which also has clear superlinear scaling, has much lower
heterogeneity than DNA-binding/transcription factors.  Among the
categories with linear scaling, Transferases is one of the least
heterogeneous ones, while the categories Protein Modification and Ion
metabolism and Transport show a large variability in the exponents of
the associated families. For Protein Modification, this signal is
essentially due to the Gro-ES superfamily and to the HFSP90 ATP-ase
domain, which have a clear superlinear scaling, while other chaperone
families, such as FKBP, HSP20-like and J-domain are clearly sublinear
with exponents close to zero. Interestingly, the Gro-EL domains,
functionally associated to the Gro-EL, are part of this second class
(exponent close to 0.2), showing very different abundance scaling to
the Gro-EL partner domains.  Conversely, the category Ion Metabolism
and Transport is divided equally into linearly scaling (e.g.,
Ferritin-like Iron homeostasis domains) and markedly sublinear
families, such as SUF (sulphur assimilation) / NIF (nitrogen fixation)
domains.  On the other hand, categories with small values of
heterogeneity are made of families with exponents close to the one of
the category, as shown in Table~\ref{tab:Table1} in the case of, e.g.,
Transferases.
% Transcription show a large variability in their family exponents
%.  In particular, the heterogeneity of
% Transcription is due to an excess of families whose sublinear scaling
% substantially differs from the category exponent (see
% Table~\ref{tab:Table1})

\subsection*{Determinants of the scaling exponent of a functional category}

We have shown that scaling exponents of individual families may
correspond to a vsriable extent to the exponent of the corresponding
functional category.
However, since categories are groups of families, the scaling of the
former cannot be independent of the scaling of the latter. This
section explores systematically the connection between the two. 
% heterogeneous scaling of families and the scaling of the associated
% categories.
%
As detailed below, we find that in some cases the scaling exponent of
functional categories is determined by few outlier families, while in
other cases most of the families within a category contribute to the
category scaling exponent.

While many families have a clear power-law scaling, functional
categories may contain many low-abundance families with unclear
scaling properties. When considered individually, these families do
not contribute much to the total number of domains of a category, but
their joined effect on the scaling of the category could be
potentially important. Supplementary Fig.~\ref{fig:low_ab_families_2}
shows that the sum of these low-abundance families does not suffer
from sampling problems and shows a clear scaling. Interestingly, the
scaling exponents for these sums once again does not necessarily
coincide with the category exponents.

\begin{figure}[htp]
\centering
\includegraphics[width=0.45\textwidth]{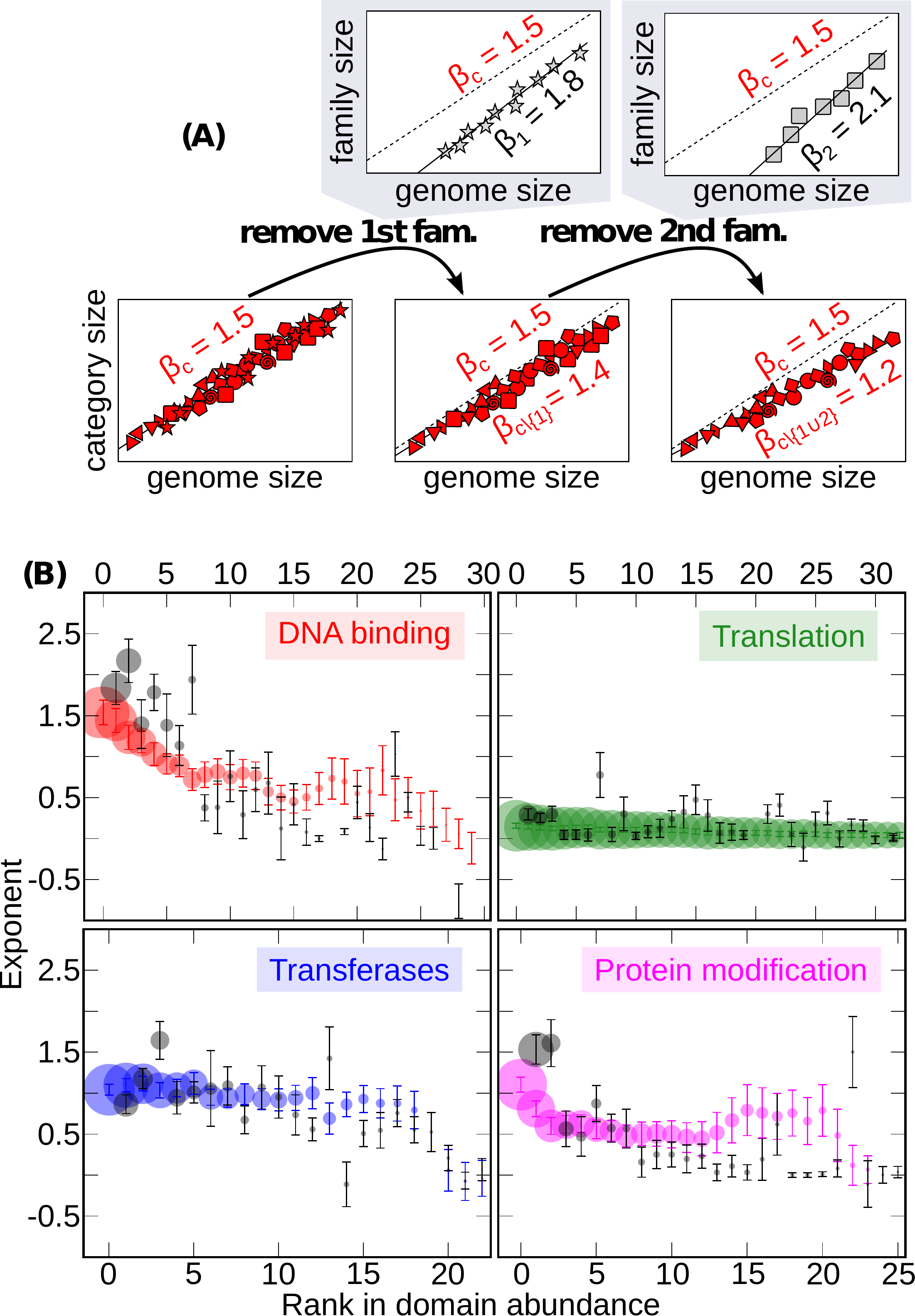}
\caption{Systematic removal of families (ranked by abundance) inside
  functional categories reveals how individual families build up
  functional category scaling. (A) Illustration of the
  procedure. Families belonging to a given functional category are
  ranked by overall abundance on all genomes and removed one by one
  from the most abundant. The scaling of the removed family and the
  remainder of the category is evaluated after each removal. The plots
  are a stylized example of the first two steps (using values for the
  category DNA binding). $\beta_{c}$ is the category exponent,
  $\beta_{i}$ are family exponents and $\beta_{c\setminus \left \{i
    \right \} }$ are the stripped-category exponents, computed after the
  removals.  (B) Results of this analysis for four functional
  categories. Grey circles represent the exponents $\beta_{i}$ (and
  their errors) for the scaling law of each family belonging to the
  functional category (in order of rank in total abundance). Colored
  circles are the scaling exponents of functional categories without
  the domains of the $i$ least abundant families. The size of each
  symbol is proportional to the fraction of domains in the family or
  family-stripped category. Error bars are uncertainties of the fits
  (see Methods). See Supplementary Fig.~\ref{fig:all_expVSrank}
  and~\ref{fig:2Bpfam} for the same plots obtained for other
  functional categories and using the PFAM database.}
\label{fig:Fig2_expVSrank}
\end{figure}

Figure~\ref{fig:Fig2_expVSrank}A illustrates the systematic procedure
that we used in order to understand how the scaling of categories
emerges from the scaling of the associated families. Families were
ranked by total abundance across all genomes (from the most to the
least abundant) and removed one by one from the category. At each
removal step in this procedure, both the scaling exponent of the
removed family and the exponent of the remainder of the category are
considered. In other words, the $i$-th step evaluates the exponent of
the $i$-th ranking family (in order of overall abundance) and of the
set of families obtained by removing the $i$ top-ranking families
(with highest abundance) from the category. The resulting exponents
quantify the contribution of each family to the global category
scaling, as well as the collective contribution of all the families
with increasingly lower overall abundance.

The results (Fig.~\ref{fig:Fig2_expVSrank}B and Supplementary
Fig.~\ref{fig:all_expVSrank} and~\ref{fig:T_expVSrank}), show how the
heterogeneity features described above are related to family
abundance.  The collective behavior of low-abundance families may
deviate sensibly from that of the functional category and families
follow scaling laws that sensibly deviate from the one of the
corresponding functional categories. One notable example of this are
Transcription-Factor DNA-binding domains.  If the abundance of the
outliers families is large enough in terms of the fraction of domains in
the functional category, they might be responsible for determining the
scaling of the entire category, as it happens in the case of
DNA-binding (which is more extensively discussed in the following
section).

Overall, one can distinguish between two main behaviors, either a
category scaling is driven by a low number of highly populated
``outlier'' families (e.g. DNA binding and Protein Modification in
Fig.~\ref{fig:Fig2_expVSrank}B), or the category scaling is coherent,
and robust to family subtraction (e.g. Transferases and Translation
in Fig.~\ref{fig:Fig2_expVSrank}B). While the first behavior appears
to be more common for functional categories with higher scaling
exponent, there are some exceptions. Notably, the scaling of strongly
super-linear categories is not always driven by a few families. For
example, the functional category Signal Transduction has an exponent
$\beta_{c}=1.7$, which remains stable after the removal of the largest
families (Supplementary Fig.~\ref{fig:low_ab_families_2}
and~\ref{fig:T_expVSrank}).  Both behaviors are clearly visible for
intermediate exponents (in order to appreciate this, compare the
Transferases and Protein Modification categories in
Fig.~\ref{fig:Fig2_expVSrank}B).

\subsection*{Super-linear scaling of transcription factors is
  determined by the behavior of a few specific highly populated
  families.}

We considered, in particular, the case of DNA-binding / transcription
factors~\cite{charoensawan2010genomic}, which are known to exhibit
peculiar scaling in bacteria~\cite{Ranea2004,Maslov2009}.  The
abundance of domains in this functional category increases
superlinearly (almost quadratically) with the total domain
counts~\cite{vanNimwegen03,Pang2011,Grilli2014}. As shown in the first
row of Fig.~\ref{fig:Fig1_fam_scaling}B, not all the families in this
functional category display a superlinear
scaling~\cite{charoensawan2010genomic}, and the collective scaling of
the low-abundance families with genome size is much slower (see
Fig.~\ref{fig:Fig2_expVSrank} and Supplementary
Fig.~\ref{fig:low_ab_families_2}).  Fig.~\ref{fig:Fig2_expVSrank}B
shows that only the most 5-6 abundant families display a super-linear
scaling ($\beta_{i}>1$). These are Winged helix DNA-binding domains
(34.8\% of abundance), Homeodomain-like (23.3 \%) lambda
repressor-like DNA-binding domains (9.5\%) bipartite Response
regulators (7.7\%) Periplasmic binding protein-like (6.2\%), and
FadR-like (2.4\%). The remaining 16.1\% of the DNA-binding regulatory
domains follows a clear \emph{sublinear} scaling with genome size
(exponent 0.7, see Supplementary Fig.~\ref{fig:low_ab_families_2}).

\subsection*{Grouping families with similar scaling exponents shows
  known associations with biological function and reveals new ones.}

The above analyses show that the range of scaling exponents of
families within the same functional categories is generally wide and
that the scaling behavior of some families sensibly deviates from
their category. At the same time, functional categories show clear
characteristic scaling laws, with well-defined exponents
$\beta_{c}$~\cite{molina2009scaling}.
We, therefore, asked to what extent a range of family scaling exponents
$\beta_{i}$ is peculiar to a functional category and how this compares
to the category exponent $\beta_c$.  To this end, we grouped families
based on their scaling exponents. We then used those groups to test
how much specific range of exponents define specific functions by an
enrichment test of functional annotations.

%~ To this end, we divided families into three groups based on their
%~ scaling exponents (families that scale linearly, super-linearly and
%~ sub-linearly). We then uses those groups to test how much specific
%~ range of exponents define specific functions by an enrichment test.
%
Table~\ref{tab:Table1} shows that in most cases functional categories
are over-represented in the exponent range where their scaling
exponents $\beta_{c}$ is found.  This confirms and puts in a wider
perspective the previously reported strong association between
abundance scaling with size and functional annotation.
As can be expected from previous results, the functional category
Protein Modification is an exception: this category is
under-represented in the linear region even though its category
exponent is $\sim 1.06$, since it contains two strongly superlinear
families and a bulk of families with sublinear scaling. This strong
heterogeneity in scaling exponents is also visible in
Fig.~\ref{fig:Fig2_expVSrank}B.
% \jac{bisognerebbe guardare lo scaling del numero di famiglie
% presenti vs famiglie totali... forse c'e' uno scaling interessante
% li' FIGURA SUPP}

The results of this analysis are not sensitive to the chosen intervals
for the scaling exponents.  In order to show this, we performed a more
systematic enrichment analysis, using sliding windows of exponents of
width 0.4, and step 0.1, and plotting the Z-score for the enrichment
as a function of the representative family exponent for each window
(Supplementary Fig.~\ref{fig:all_bin}).  The maxima of this plot
define a representative exponent for each functional category, and can
be compared to the exponent $\beta_c$ measured directly from the plot
of category abundance vs genome size (see Supplementary Figure
\ref{fig:realExp_maxZexp}). Interestingly, this analysis also shows
that in many cases a single functional category is enriched for
multiple groups of families with well-defined exponents, as in the
case of the Protein Modification category. The cases of Ion Metabolism
and Transport (already discussed), Coenzyme Metabolism and Transport,
Redox also shows clear indications of enrichment for two or more
exponent groups. For the category Coenzyme Metabolism and Transport
this is due to the presence of a single abundant family with scaling
exponent close to 2, the acyl-CoA dehydrogenase NM domain-like, whose
functional annotation is still not well defined. In the case of Redox,
the most abundant families (Thioredoxin-like, 4Fe-4S ferredoxins,
Metallo-hydrolase/Oxydoreductase) scale linearly, but there is a wide
range of families with exponents between 0.5 and 1, and once again two
fairly abundant outlier families with superlinear scaling
(Glyoxalase/Bleomycin resistance protein/Dioxygenase, and ALDH-like),
both with a fairly wide range of functional annotations.

\begin{table}[ht!!]
\centering
\begin{tabular}{c|l|c|c|c|c}
\hline
\hline
\rule{0pt}{2.5ex}
Detailed function		&	$\beta_i \leq 0.6$	&	$0.6 < \beta_i < 1.4$	&	$\beta_i \geq 1.4$	&	$\beta_{c}\pm\sigma_{\beta_{c}}$	\\ [0.5ex]
\hline											
Translation				&	20(\ooverrep{$4.3$})	&	1(\uunder{$-3.7$})		&	0					&	$0.16 \pm 0.03$	\\ [0.5ex]
DNA replication/repair	&	11						&	7						&	0					&	$0.51 \pm 0.07$	\\ [0.5ex]
\hline																
Transport				&	5						&	9						&	1					&	$1.1 \pm 0.2$	\\ [0.5ex]
Proteases				&	7						&	9						&	0					&	$0.9 \pm 0.1$	\\ [0.5ex]
Protein modification	&	8 						&	1(\uunder{$-2.3$})		&	2					&	$1.06 \pm 0.09$	\\ [0.5ex]
Ion m/tr 				&	11						&	3						&	3(\uunder{$-2.2$})	&	$1.3 \pm 0.1$	\\ [0.5ex]
\hline											
Other enzymes			&	29						&	32						&	2					&	$1.04 \pm 0.06$	\\ [0.5ex]
Coenzyme m/tr			&	17(\ooverrep{$2.2$})	&	6						&	1					&	$0.85 \pm 0.09$	\\ [0.5ex]
Redox					&	4(\uunder{$-3.3$})		&	18(\ooverrep{$3.1$})	&	2					&	$1.2 \pm 0.1$	\\ [0.5ex]
Energy					&	11						&	7						&	0					&	$0.86 \pm 0.09$	\\ [0.5ex]
Nucleotide m/tr			&	16(\ooverrep{$3.1$})	&	3(\uunder{$-2.5$})		&	0					&	$0.53 \pm 0.08$	\\ [0.5ex]
Carbohydrate m/tr		&	4						&	8						&	0					&	$1.0 \pm 0.2$	\\ [0.5ex]
Transferases 			&	5						&	11						&	1					&	$1.05 \pm 0.07$	\\ [0.5ex]
Amino acids m/tr		&	7						&	6						&	0					&	$0.8 \pm 0.2$	\\ [0.5ex]
\hline																					
DNA-binding				&	5						&	4						&	4(\ooverrep{$3.3$})	&	$1.5 \pm 0.1$	\\ [0.5ex]
Signal transduction		&	1(\uunder{$-2.7$})		&	5						&	5(\ooverrep{$5.0$})	&	$1.6 \pm 0.2$	\\ [0.5ex]
\hline													
Unknown function	 	&	9						&	7						&	0					&	$0.98 \pm 0.09$	\\ [0.5ex]
\bottomrule
\end{tabular}
\caption{Family scaling exponents can be associated to specific
  biological functions. Each 
  cell in the table indicates the number of families that 
  functional categories (rows) share with groups of families whose
  scaling exponents  fall  in pre-defined intervals
  (columns). The table also shows the Z-scores
  for a standard hypergeometric test  (shown in green for
  over-representation and in red for under-representation, only
  $|Z|>$1.96 are shown). }
\label{tab:Table1}
\end{table}

%~ 
%~ These deviations may be compared with the typical fluctuations of
%~ $\beta_i$ obtained randomizing the data accordingly to the null model
%~ described above. However, as we already discussed, these fluctutions
%~ are very small in the null model, so the comparison is not meaningful.

% Interestingly, the exponent of the family that deviates most from
% the null model in each category is linearly related with the
% category exponent.  Thus, the typical fluctuations of family
% exponents are related the scaling exponent of the category.  This
% result can be simply understood in terms of the constraints imposed
% by the category exponent to the scaling of single families.
% Categories with very small exponents are incompatible with large
% fluctuations of family exponents.

\section{DISCUSSION AND CONCLUSIONS}

Our results gather a critical mass of evidence in the direction of
family-specific expansion rules for the families of protein domains
found in a genome.
Although previous work had focused on individual
transcription factor families~\cite{charoensawan2010genomic}, finding
in some cases some definite scaling, no attempts were made to address
this question systematically.
The scaling laws for domain families appear to be very robust, despite
of the limited sampling of families compared to functional annotations
(which are super-aggregates of families and hence have by definition
higher abundance). In particular, the results are consistent between
the different classifications of families we tested (SUPERFAMILY and
PFAM, see SI, sec.~S4).
%Importantly, a null model shuffling the family abundance table 

%
Overall, our results indicate that scaling laws are measurable at the
family level, and, given the heterogeneous scaling of families with
the same functional annotations, families are likely a more reliable
description level for the scaling laws than functional annotations.
The interpretation of these scaling laws is related to the
evolutionary dynamics of family expansion by horizontal transfer or
gene duplication, and gene
loss~\cite{Koonin2011,vanNimwegen03,Grassi2012}.
Scaling exponents are seen as ``evolutionary
potentials''~\cite{Molina08}, is based on a model of function-specific
(multiplicative) family expansion rates.  Assuming this
interpretation, then our result that these rates may be different for
different domain families having the same functional annotation may
seem puzzling.
Clearly, selective pressure can only act at the functional level, and
if two folds were functionally identical, there should be reasonably no
advantage selecting one with respect to the other, and doing so at
different specific rates. For example, a transcription factor using
one fold to bind DNA rather than another one should be
indistinguishable from one using a different fold, provided binding
specificity and regulatory action are the same.

In view of these considerations, we believe that our findings support
a more complex scenario for the interplay between domain families and
their functions.
Specifically, we put forward two complementary rationalizations, both
of which are probably in part verified in the data. The first is that
functional annotations group together different domains whose
abundance is linked in different ways to genome size because of their
different biochemical and biological functional roles. Such differences
may range from slight biochemical specificities of different folds to
plain misannotations.
This is possible, e.g., with enzymes, where the biochemical range of
two different folds is generally different. This obervation might be
related to the positive correlation we found between the number of EC
numbers corresponding to a metabolic domain and its scaling exponent.
However, such interpretation might be less likely applicable to, e.g.,
transcription factor DNA-binding domains, where functional annotation
is fairly straightforward~\cite{Babu03}, but different scaling
behaviors with genome size are nevertheless found. 

The second potential explanation assumes the point of view where
scaling laws are the result of functional interdependency between
different domain families~\cite{Maslov2009,Grassi2012a}, then
correlated fluctuations around the mean of family pairs should carry
memory of such dependency structures~\cite{Pang2013}.
More in detail, there may be specific dependencies connecting the
relative proportions of domains with both different and equal
functional annotations that are present in the same genome, which
might determine the family-specific behavior~\cite{Grilli2012}. While
further analysis is required to elucidate these trends, we believe
that gaining knowledge on functional dependencies would be an
important step to understand the functional design principles of
genomes.
%

%%%%

Of notable importance is the case of the superlinear scaling of
transcription factors, which has created notable debate in the
past~\cite{Ranea2005,Maslov2009}.  For the first time, we look into
how this trend is subdivided between the different
DBDs~\cite{Babu03}. Our analysis indicates that the superlinear
scaling is driven by the few most abundant superfamilies (mostly
winged-helix, homeodomain, lambda repressor). However, the remaining
10-20\% of the functional category gives a clear sublinear scaling with
genome size, which emerges beyond any sampling problems. We speculate
that these other regulatory DNA-binding domains may be functionally
different or behave differently over evolutionary time scales.
Hence, the scaling of transcription factors with size in bacteria is
driven by a small set of domain families with scaling exponent close
to two, which take up most of the abundance, but does not appear to be
peculiar of \emph{all} transcription factors.
A ``toolbox'' model considering the role of transcription factors as
regulator of metabolic pathways and the finite universe of metabolic
reactions~\cite{Maslov2009,Pang2011} predicts scaling exponents close
to two for transcription factor families. According to our results,
such model should be applicable to the leading TF families.
Interestingly, the heterogeneity in the behavior in transcription
factor DNA-binding domains is much higher than that of the other
notable superlinear functional category, signal transduction, where
removal of the leading families does not significantly affect the
observed scaling of abundance with genome size. Given the clarity and
uniformity of the scaling exponent, we speculate that possibly a
toolbox-like model may be applicable to understand the overall scaling
of this category.

%%% \textbf{qua in realta' sarebbe interessante vedere eucarioti... }

%%% anche cosa degli EC number e ripescaggio del coding tlusty alon

%%%%%

Other categories clearly contain multiple sets of families with
coherent exponents or single outlier families. In some cases, two main
groups of families with different scaling behavior clearly emerge, and
higher observed scaling exponents may be related to a wider range of
functional annotations. 
We propose that such easily detectable trends can be used to revise
and refine functional annotations of protein domains. Such functional
annotations are currently largely curated by humans, and based on
subjective and/or biased criteria. The analysis of family scaling
gives an additional objective test to define the coherence of the
families that are annotated under the same function.
While yet-to-be-developed automated inference methods based on our
observations could serve this purpose, the quantitative scores defined
here already provide useful information. The heterogeneity of a
functional category is an indication of how likely  that group of
domain families follows a coherent expansion rate over evolution. 
The enrichment scores for sets of families with a given range of
scaling exponent helps to pinpoint the sets of families within the
functional category that expand coherently with genome size.

\section{ACKNOWLEDGEMENTS}
We thank Erik van Nimwegen, Madan Babu, Otto Cordero and Purushottam
Dixit for helpful discussions.

\newpage

\bibliographystyle{nar} % Style BST file
\bibliography{scalingbibs}

\newpage
\clearpage

\setcounter{figure}{0}
\setcounter{section}{0}
\renewcommand{\figurename}{Supplementary Figure}
\renewcommand{\thefigure}{S\arabic{figure}}

\renewcommand{\thesection}{S\arabic{section}}
\setcounter{table}{0}
\renewcommand{\tablename}{Supplementary Table}
\renewcommand{\thetable}{S\arabic{table}}

\begin{center}
 
  \textbf{\Large Supplementary Material for De Lazzari \emph{et al.} \\
    The scaling laws in the bacterial genomic repertoires are
  family-specific. } \\

  \vskip1.3cm

  Eleonora de Lazzari, Jacopo Grilli, Sergei Maslov and Marco
  {Cosentino~Lagomarsino}

  \vskip1.3cm

\end{center}

\section*{Supplementary Notes}

\section{Filters for reliable family scaling}
\label{sec:famthresh}

Many domain families are found only in a few genomes and/or in very
few copies. For this reason, they might not show clear scaling
properties.  We excluded such families from the analysis with some
filtering criteria.  In order to filter out these families, we used
three independent parameters.  The first one is the occurrence of a
family (i.e., the fraction of genomes where it is present, and the
number of points available for the fit), the second one is the Pearson
correlation of its abundance with genome size and the third one is the
goodness of fit to a power law.

The occurence of the family $i$ is defined as
\begin{equation}
o_{i} = \frac{ N_{G}^{(i)} }{ N_{G} } \ ,
\end{equation}
while $N_{G}^{(i)}$ is the number of genomes in which family $i$ is
present and $N_{G}$ is the total number of genomes in the sample.

\begin{figure}[!htbp]
\centering
\includegraphics[width=0.8\textwidth]{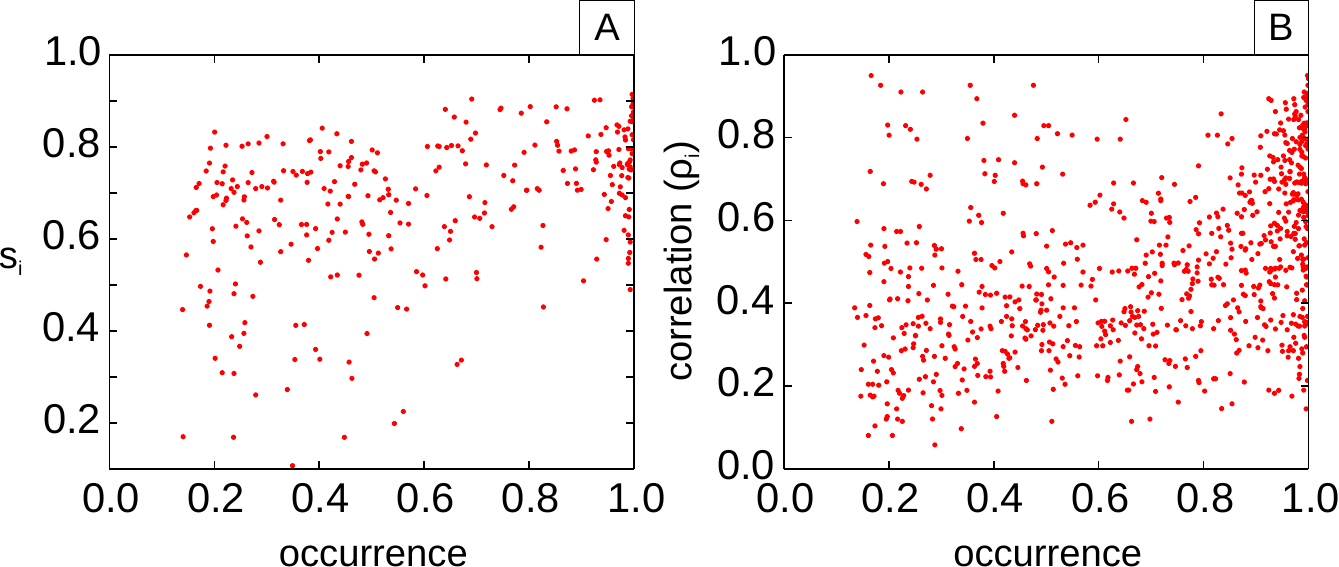}
\caption{ \textbf{The parameters used to filter out families are
    independent.}  (A) The plot reports the goodness of fit index
  $s_i$, which is the average squared deviation between the empirical
  family abundance and the one derived from the fit, as a function of
  family occurrence. Each point represents a family whose exponent is
  lower than 0.2 .  (B) Pearson correlation between family abundance
  (number of domain belonging to a given family) and genome size,
  calculated across the genomes where the family is present as a
  function of family occurrence. Each point represents a family whose
  exponent is higher than 0.2 .  The lack of clear correlation visible
  in the plots shows that the three indices are all relevant in the
  filters.
%     Family size - genome size correlation and the goodness of
%     fit index are uncorrelated with family occurrence. In both panels,
%     the $x$-axis displays the family occurrence, i.e. the fraction of
%     genomes in which a given family is present. (A) The $y$-axis shows
%     the goodness of fit index $s_i$, which is the average squared
%     deviation between the empirical family abundance and the one
%     derived from the fit. Each point represents a family whose
%     exponent is lower than 0.2.  (B) The $y$-axis shows the value of the
%     Pearson correlation between family abundance (number of domain
%     belonging to a given family) and genome size, calculated across
%     the genomes where the family is present. The lack of clear
%     correlation visible in the plots shows that the three indices are
%     all relevant in the filters.
}
\label{fig:corroccurence}
\end{figure}

The Pearson correlation coefficient $\rho_{i}$ between the logarithm
of the family abundance and the logarithm of the genome size
quantifies the existence of a relation between family abundance and
genome size.  It should be noted that this quantity differs from
$o_i$: these two properties are not correlated in the data
(Fig.~\ref{fig:corroccurence}B), which indicates that filters on their
value should be applied independently.

Considering families with clear but shallow scaling or constant
abundance across genomes, the Pearson correlation index gives values
close to zero, or slighlty negative, therefore another parameter is
required to assess the accuracy of the fit results.  For each family
whose estimated scaling exponent is lower than 0.2, we defined the
quantity $LS_{i}$ as
\begin{displaymath}
  LS_{i} = \dfrac{1}{N_{G}^{(i)}}\sum\limits_{g=1}^{N_{G}^{(i)}} 
  \dfrac{\left[ 
 y_{fit, i} - y_{i}  
  \right]^2}{ y_{fit} }  \ , 
\end{displaymath}
where $y_{i}$ is the logarithm of the empirical abundance of family
$i$ and $y_{fit, i}$ is the abundance calculated with the fit
parameters, i.e.,
\begin{displaymath}
  y_{fit, i} = \alpha_{i} + \beta_{i}\log\left(\sum\limits_{i=1}^{N_F}
    n_{i}^g \right)  \ . 
\end{displaymath}
The goodness of fit index $s_i$ was defined as
\begin{displaymath}
s_{i} = \dfrac{1}{1+\sqrt{LS_{i}}}	
\end{displaymath}
so that the values of $s_i$ close to 1 correspond to the minimum value
of the average squared deviation between the fit and the empirical
values.  The goodness of fit index $s_i$ is independent from the
occurrence $o_{i}$ as shown in Fig.~\ref{fig:corroccurence}A.

\begin{figure}[!htbp]
\centering
\includegraphics[width=0.8\textwidth]{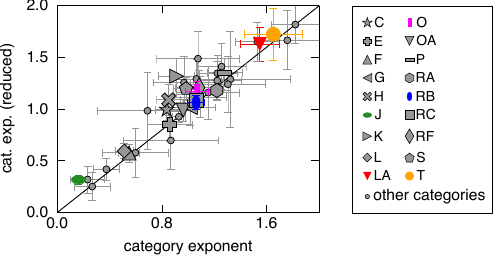}
\caption{\textbf{Category exponents are robust by removal of filtered
    families}. The plot compares the category exponent obtained by
  considering all the domains and the exponent obtained by removing
  from the category count the domains belonging to families filtered
  out by our criteria for unclear scaling.  The exponents before
  ($x$-axis) and after thresholding ($y$-axis) are compatible within
  their errors. The solid line is the $y=x$ line. The panel on the
  right shows the association between symbols and category codes (see
  Table~\ref{tab:TableS0sym} for the corresponding category name).  }
\label{fig:cat_cfr}
\end{figure}

We considered only the families with $o_i > 0.6$.  If the fitted
scaling exponent is higher than $0.2$ then we excluded the families
with $\rho_i < 0.4$, while, for exponents lower than $0.2$, only
families with $ s_i > 0.9 $ were taken into account. After this
thresholding, we removed 1179 families out of 1536. While the fraction
of such small and sparse families is large, we verified that they do
not contribute significantly to the category scaling (see
Figure~\ref{fig:cat_cfr}).

\section{Formulation of the null model for family-specific scaling}
\label{sec:nullmodel}

\begin{figure}[!htbp]
\centering
  \includegraphics[width=0.45\textwidth]{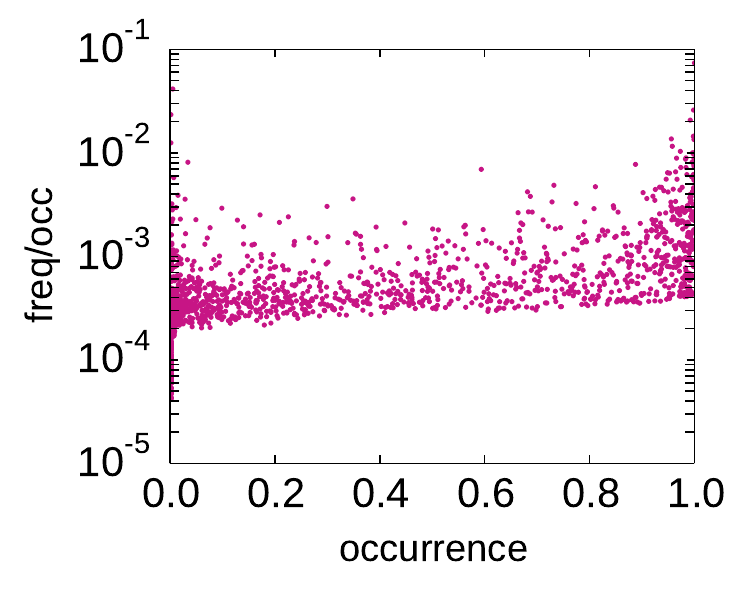}
  \caption{\textbf{Absence of correlation between family frequency and
      occurrence in empirical data}. Ratio between the family
    frequency and family occurrence plotted vs family occurrence. The
    frequency of a family in a genome is defined as the ratio between
    its abundance and the genome size in domains, i.e. the total
    number of domains found on the genome. The plot shows that there
    is no clear correlation between the two quantities, implying that
    universally found (``core'') families are not necessarily more
    abundant than rare families, and viceversa. Note that, given a
    value of occurence, there is a technical lower bound to the
    frequency. If a family is present in $G$ genomes, the mimimum
    value that the frequency can assume is when the family is present
    with only one domain in the $G$ \emph{largest} genomes, and
    therefore have frequency lower than the inverse of the size of the
    $G$-th largest genome.  It is expected that the lower bound
    increases with occurrence.}
\label{fig:occvsfreq}
\end{figure}

The null model assumes that the total abundance of a category is
distributed randomly across the families belonging to it. Both the
average relative abundance (i.e. fraction of domains belonging to a
family averaged over genomes) and occurrence (fraction of genomes
where the family is present) of each family are conserved (they are in
fact two independent properties, see Fig.~\ref{fig:occvsfreq}).

The null model is based on the following ingredients.
\begin{itemize}
\item The number of domains belonging to a category $c$ in genome $g$,
  $n_{i}^{g}$, is conserved.
\item For each genome, domains are not assigned to families that are
  not present in that genome.
\item The average frequency for each family with respect to the
  category is conserved.
\end{itemize}
The average frequency $f_c(i)$ of the family $i$ with respect to the
category $c$ is defined as
\begin{equation}
  f_c(i) = \frac{1}{N_{G}^{(i)}} \sum_g \dfrac{n_{i}^{g} }{ n^{g}_c } \ ,
\end{equation}
where the family index $i$ belongs to the set in category $c$ and the
sum over $g$ is carried over all the genomes, while $N_{G}^{(i)}$ is
the number of genomes in which family $i$ is present, $n_{i}^{g} $ is
the abundance of family $i$ in genome $g$ and $n^{g}_c = \sum_{i \in
  c} n_{i}^{g}$ is the abundance of category $c$ in genome $g$.

Given a genome $g$, each realization of the null model redistributes
randomly the $n^{g}_c$ domains of the functional category $c$ arranged
in the $F^g_c$ families belonging to category $c$ in genome $g$.  Each
one of the $n^{g}_c$ domains is assigned to family $i$ with
probability
\begin{equation}
\displaystyle
p_{c}(i) =
\begin{cases}
\displaystyle
  \frac{f_c(i)}{\sum_{i \in c}f_c(i)}  , & \mbox{if }n_{i}^{g}\neq
  0 \\
  0, & \mbox{if }n_{i}^g=0 \ .
\end{cases}
\end{equation}

\section{Enrichment analysis for functional categories and families
  with similar scaling exponent}
\label{sec:grouptest}

All families passing the filters described in section
\ref{sec:famthresh} were divided into three groups based on the values
of their exponent $\beta_{i}$:
\begin{itemize}
\item sub-linearly scaling families, $\beta_i \leq 0.6$
\item linearly scaling families, $0.6 < \beta_i < 1.4$
\item super-linearly scaling families, $\beta_{c} \geq 1.4$
\end{itemize}
We used hypergeometric tests to asses over- or under-representation of
functional categories in these family groups. Given that $F_{c}$ is
the number of families that belong to the category $c$, $F_{bin}$ is
the number of families in either of the three groups defined above and
$F_{tot}$ is the total number of families involved in this analysis,
the mean and the variance of the hypergeometric distribution are:
\begin{displaymath}
\mu_{bin, c} =  F_{bin}  \cdot  \dfrac{F_{c} }{F_{tot}}  \ ,
\end{displaymath}
\begin{displaymath}
  \sigma_{bin, c}^2 = F_{bin} \cdot  \dfrac{ F_{c} }{F_{tot}}  \cdot \left( 1 - \dfrac{ F_{c} }{F_{tot}} \right) \cdot \left( 1 - \dfrac{F_{bin}-1}{F_{tot}-1} \right)  \ ,
\end{displaymath}
For each combination of functional category and family group, the
quantity $x_{bin, c}$ is the number of families that lie in the
intersection of category $c$ with family group $bin$. The functional
category $c$ is under-represented in the group $bin$ if $Z_{bin,
  c}<-1.96$, over-represented if $Z_{bin, c}>-1.96$, where $Z_{bin,c}$
is the Z-score:
\begin{displaymath}
Z_{bin, c} = \dfrac{ x_{bin, c} - \mu_{bin,c}}{\sigma_{bin, c}}  \ ,
\end{displaymath}
The resulting intersection values $x_{bin, c}$ and the significant
Z-scores are reported in Table~1 of the main text.

In order to prove that the results are independent from the chosen
interval of the exponents, we substituted the three groups with
sliding intervals of amplitude 0.4 and step 0.1 and repeated the same
process.  Only intervals with more than 10 families are considered.
Fig.~\ref{fig:all_bin} shows that the results are consistent with the
previous analysis.  Fig.~\ref{fig:realExp_maxZexp} shows how the
exponent corresponding to the maximum Z-score differs in some cases
from the category exponent.

\section{Main results hold also for PFAM clans}
\label{sec:pfam}

We chose PFAM clans as an alternative database to test our results.
PFAM clans were annotated on the same scheme of 50 functional
categories used for superfamilies, using the mapping of clans into
superfamilies available from the PFAM website
\url{http://pfam.xfam.org/clan/browse#numbers}~\cite{finn2006pfam}.
The scaling laws for functional categories are recovered also for
clans (Fig.~\ref{fig:sfVSclan} and Table~\ref{tab:TableS1cat}) and are
consistent with previous
results~\cite{vanNimwegen03,Molina08,MOLINA09,cordero2009regulome,Grilli2012,charoensawan2010genomic}.

The following main results were recovered for Pfam clans.
\begin{itemize}
\item For each clan, the abundance across genomes scaleS as a power law
  of the genome size. Equally to SCOP superfamilies, Pfam clans have
  individual scaling exponents that may or may not follow the one of
  the associated functional category (Table.~\ref{tab:TableS2clan}).
  The fitting method and threshold values are the same used for
  supefamilies (sec.~\ref{sec:famthresh}).  178 clans out of 446
  passed the filters and were employed for further analysis.
\item The heterogeneity (average of the distance between the category
  exponent and the clan exponent), positively correlates with the
  category exponent (Fig.~\ref{fig:HcVSexp_pfam}). Functional
  categories with superlinear scaling tend to be more heterogeneous
  and, as found for superfamilies, the functional category Signal
  Transduction is less heterogeneous than DNA-binding, although having
  the largest exponent.  Unlike the case of superfamilies, Protein
  Modification does not have high heterogeneity score, but the
  difference in scaling between the (strongly superlinear) outlier
  family Gro-ES and the remaining ones is observed.  For clans, the
  scaling of Protein Modification is once again strongly biased by
  the clan ``GroES-like superfamily'' (20\% of the total domains).
\item Either few or most of the clans determine the scaling exponent
  of the functional category they belong to. Figure~\ref{fig:2Bpfam}
  is coherent with what observed for superfamilies, in particular the
  functional category of DNA-binding is dominated by one clan (the
  ``Helix-turn-helix'' clan) that accounts for 83\% of the total
  domains. As for superfamilies, Signal Trasduction is robust to the
  progressive removal of families confirming that the presence of
  dominant clans is not related to the superlinear scaling of the
  category.
\item Grouping clans with similar scaling exponents recovers known
  associations between the category exponent and the biological
  function (Fig.~\ref{fig:realExp_maxZexp_pfam}).
\end{itemize}

\section{Correlation between family scaling and EC numbers}
\label{sec:ECnum}
The Enzyme Commission (EC) number is a classification scheme for
enzyme-catalyzed chemical reactions. It is built as a four-levels tree
where the top nodes are six main groups of reactions, namely
Oxidoreductases, Transferases, Hydrolases, Lyases, Isomerases and
Ligases.
We used the mapping between Superfamilies and EC terms~\cite{Gough01},
to investigate the correlation between the Superfamily scaling and the
number of different reactions in which the family is involved. This
quantity is the count of distinct EC numbers corresponding to the
finest level of the EC classification.
This number shows a positive correlation (Spearman 0.74) with the
scaling exponent of metabolic family (see Table~\ref{tab:TableCorrelations}). The diversity of EC numbers in
metabolic families is also correlated with the mean total abundance of
a family, since family abundance and scaling exponent are also
correlated (Spearman 0.72).

% number of distinct EC numbers as a proxy to the variety of functions
% related to one family

%%%%%%%%%%%%%%%%%%%%%%%%%%%%%%%%%%%%%%%%%%%%%%%%%%%%%%%%%%%%%%%%%%%%%%%%%%%

\bibliography{../scalingbibs}

\clearpage

\newpage

\section*{Supplementary Figures}

%%% FIGURA S1A - 'Straightness' vs Occorrenza
%%% FIGURA S1B - 'Correlation' vs Occorrenza

\clearpage
\newpage
%%% FIGURA S2 - Exp before and after threshold

\clearpage
\newpage
%%% FIGURA S3 - Occorrenza vs Frequenza

\clearpage
\newpage
%%% FIGURA S4 - Null model scaling
\begin{figure}[!htbp]
\centering
  \includegraphics[width=0.8\textwidth]{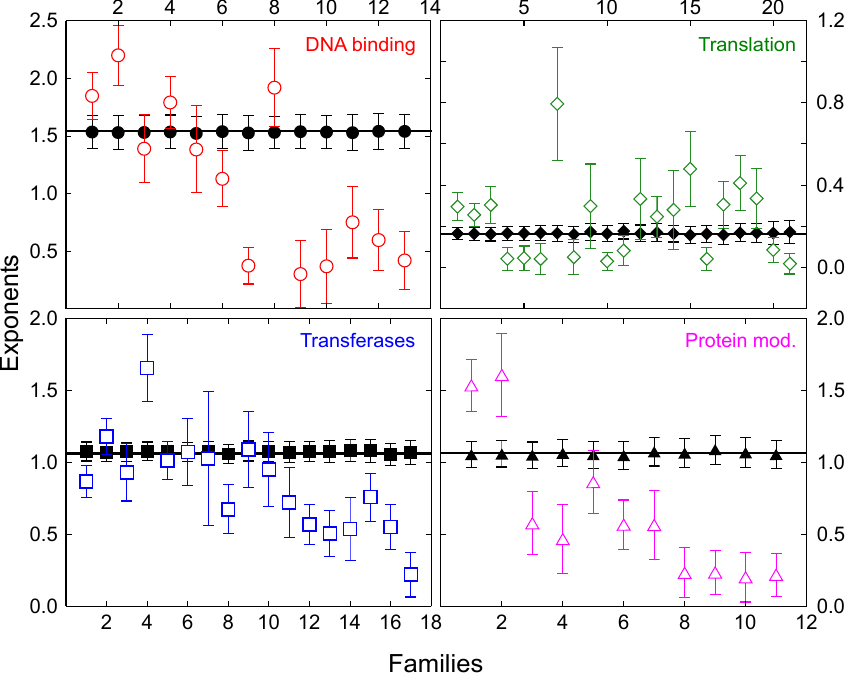}
  \caption{\textbf{Family exponents differ significantly from the null
      expectation set by the scaling of the associated functional
      category}. In order to account for random fluctuations in family
    composition within a category, the family-exponents (colored
    symbols with error bars) were compared with the ones calculated
    randomizing the data accordingly to the null model presented in
    section~\ref{sec:nullmodel} (black squares, error bars are
    variability across 1000 realizations). The variability obtained
    from the null model is extremely low and is not sufficient to
    explain the variability of scaling exponents of different families
    within a category. Each panel corresponds to a different functional
    category, its scaling exponent is shown as the black horizontal
    line. Families within each category are sorted in decreasing order
    of abundance, i.e. total domain count in the category. }
\label{fig:nullmodel}
\end{figure}

\clearpage
\newpage
%%% FIGURA S5 - Scaling of the sum of low abundance families
\begin{figure}[!htbp]
\centering
\includegraphics[width=0.55\textwidth]{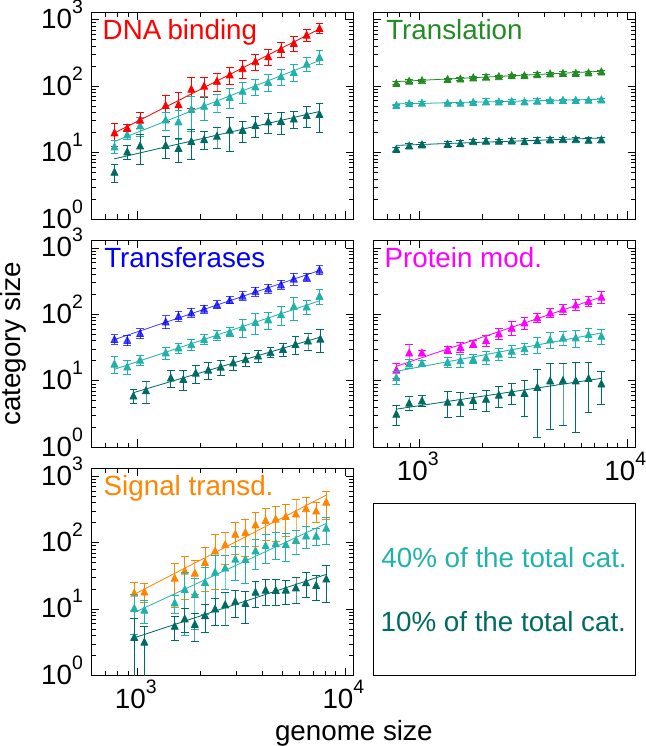}
\caption{\textbf{Scaling of the abundance of domains belonging to the
    least abundant families.} We progressively removed families from
  the category in decreasing order of family abundance and represented
  the scaling of the abundance of the remaining stripped category
  ($y$-axis) with genome size ($x$-axis). The scaling of the stripped
  category with $\approx 40\%$ initial abundance is shown in light
  green, while the category with $\approx 10\%$ of the initial
  abundance in dark green.  The scaling of the original category is
  shown with a different color (category-specific, as in Fig.~1 and~2
  of the Main Text) in each panel.  }
\label{fig:low_ab_families_2}
\end{figure}

\clearpage
\newpage
%%% FIGURA S6 - Fig 3B fatta per tutte le categories di SUPFAM
\begin{figure}[!htbp]
\centering
  \includegraphics[width=0.95\textwidth]{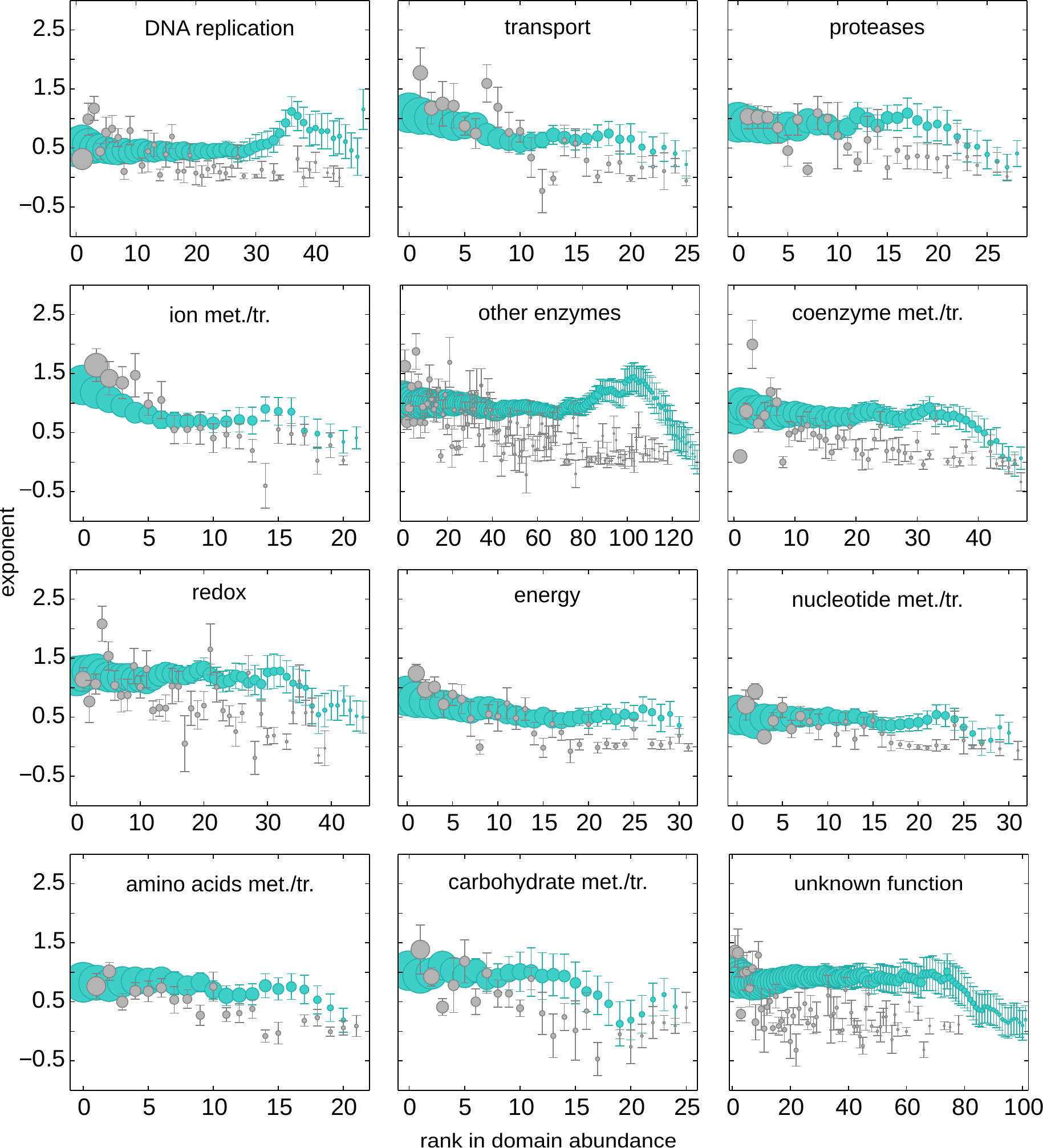}
  \caption{\textbf{Systematic removal procedure showing the role of family
      abundance in building the scaling laws of functional
      categories.} Same as Fig.~3B of the main text, for all the
    categories. Grey circles represent the exponents $\beta_i$ (and
    their errors as error bars) for the scaling law of each family
    belonging to the functional category (in order of rank in total
    abundance). Cyan circles are the scaling exponents of stripped
    functional categories, without the domains of the $i$ most
    abundant families. The size of each symbol is proportional to the
    fraction of domains in the family or family-stripped
    category. Error bars are uncertainties of the fits (see Methods).}
\label{fig:all_expVSrank}
\end{figure}

\clearpage
\newpage
%%% FIGURA S7 - Fig 3B fatta SIGNAL TRANSDUCTION
\begin{figure}[!htbp]
\centering
  \includegraphics[width=0.4\textwidth]{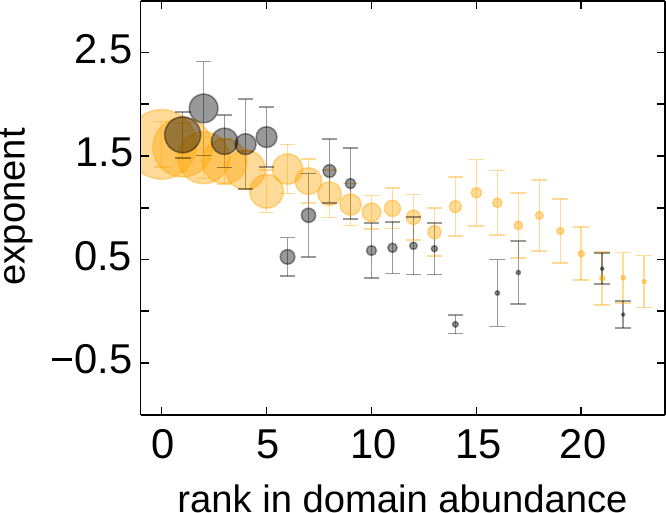}
  \caption{\textbf{The Signal Transduction funcitonal category shows
      the most coherent superlinear scaling, with exponent close to
      2}. Same as Fig. 3B of the main text for the functional category
    Signal Transduction. Grey circles represent the exponents
    $\beta_i$ (and their errors) for the scaling law of each family
    belonging to the functional category (in order of rank in total
    abundance). Orange circles are the scaling exponents of Signal
    Transduction without the domains of the $i$ most abundant
    families. The size of each symbol is proportional to the fraction
    of domains in the family or family-stripped category. Error bars
    are uncertainties of the fits (see Methods).}
\label{fig:T_expVSrank}
\end{figure}

\clearpage
\newpage
%%% FIGURA S8 - Fig 3B fatta per PFAM
\begin{figure}[!htbp]
\centering
  \includegraphics[width=1\textwidth]{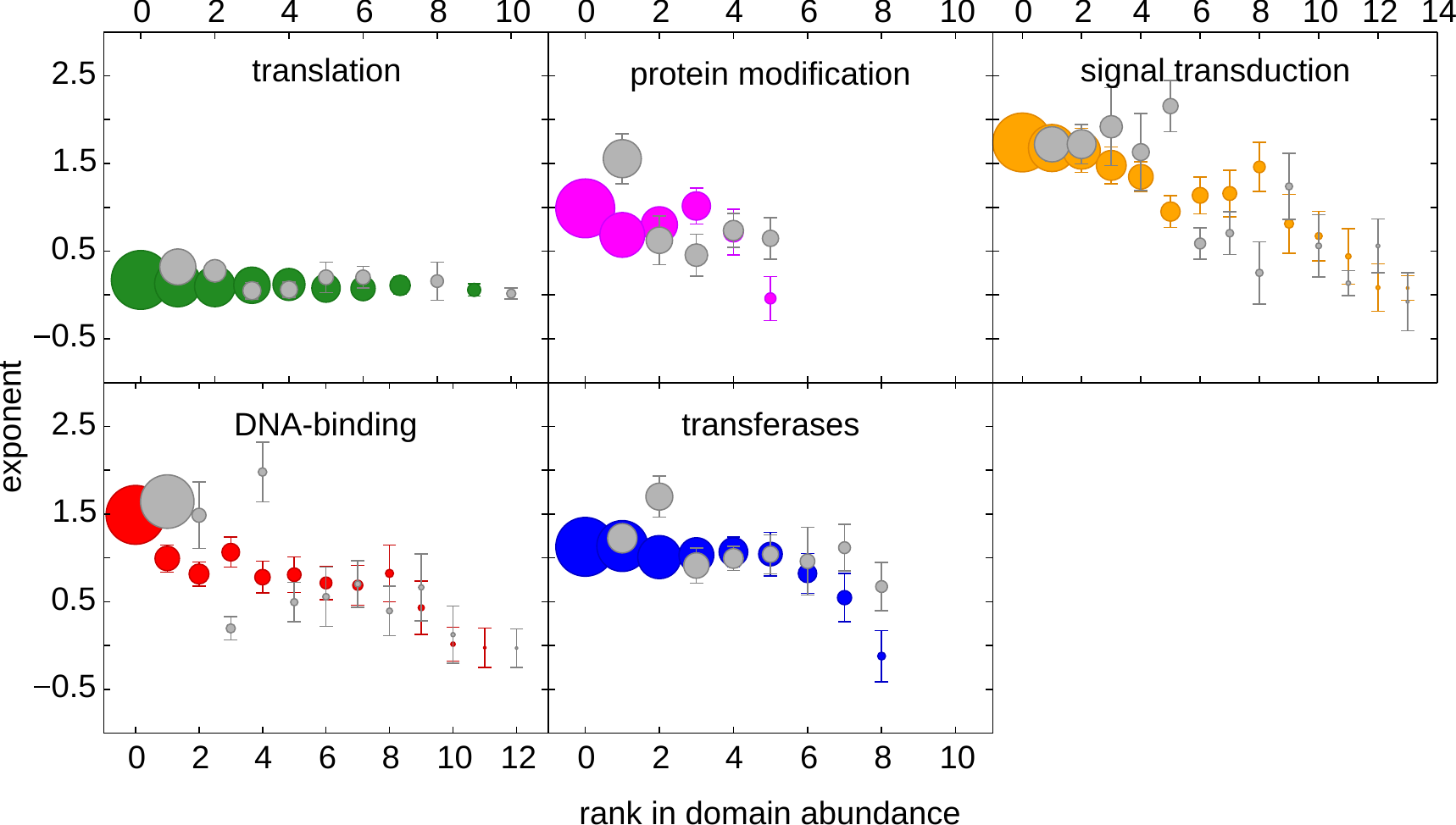}
  \caption{\textbf{Systematic removal of families (ranked by
      abundance) from functional categories reveals how individual
      families build up functional category scaling.} Same as Figure
    3B of the main text and Figure~\ref{fig:all_expVSrank}, for Pfam
    clans. Grey circles represent the exponents $\beta_i$ (and their
    errors) for the scaling law of each clan (instead of SUPFAM
    families) belonging to the functional category (in order of rank
    in total abundance). Colored circles are the scaling exponents of
    functional categories without the domains of the $i$ most abundant
    clans. The size of each symbol is proportional to the fraction of
    domains in the clan or clan-stripped category. Error bars are
    uncertainties of the fits (see Methods).}
\label{fig:2Bpfam}
\end{figure}

\clearpage
\newpage
%%% FIGURA S9- Z-scores when families are grouped into sliding bins
%%% for all categories
\begin{figure}[!htbp]
\centering
  \includegraphics[width=0.6\textwidth]{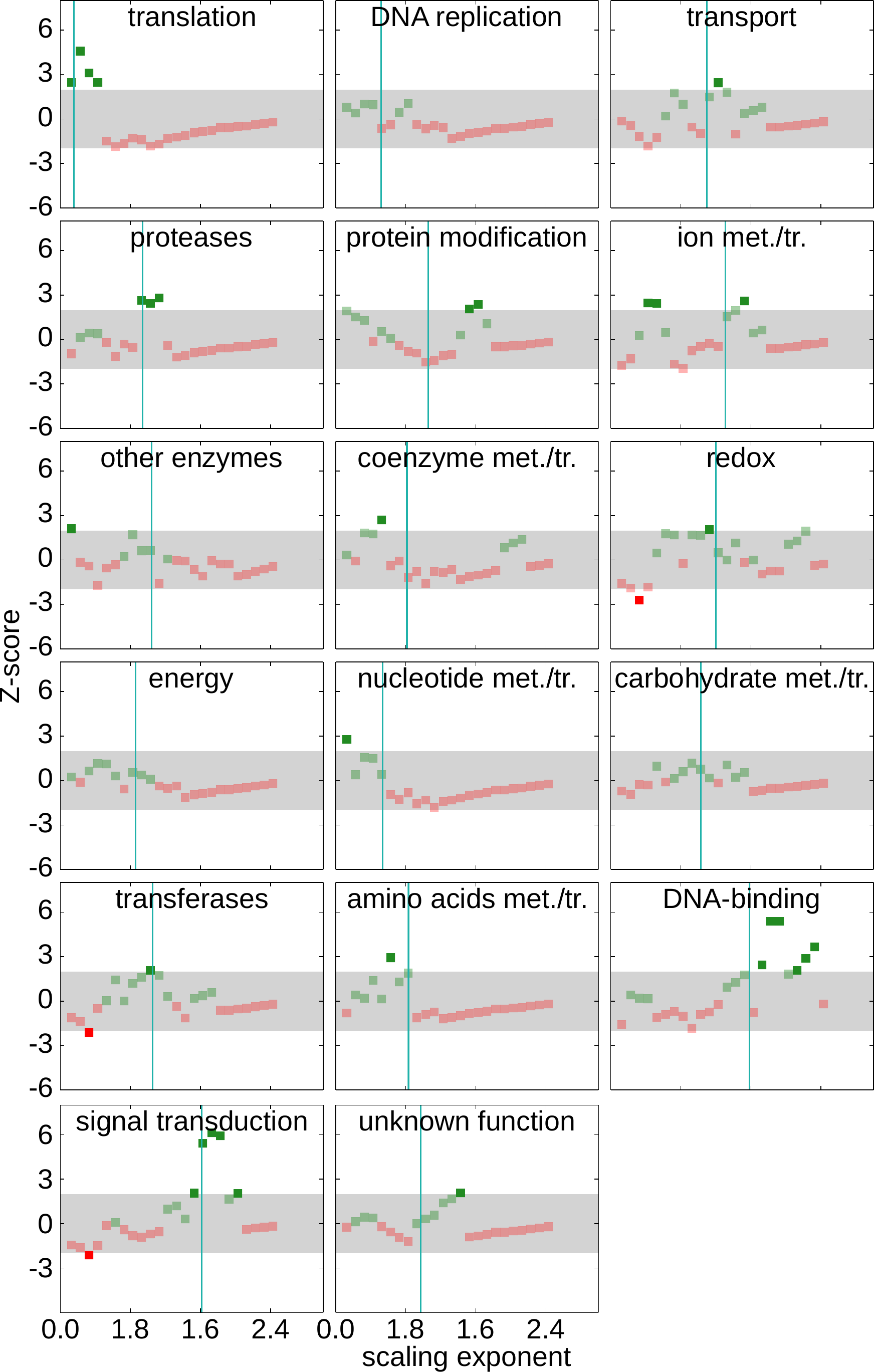}
  \caption{\textbf{A functional enrichment test for sets of families with
      similar scaling exponents}.  Families are grouped into sliding bins
    according to the value of their scaling exponent and tested for
    enrichment against each functional category. Each panel shows the
    results of the enrichment test for all functional categories with
    more than 10 families. The $x$-axis represents the center of the
    interval of exponents defining the family set. The $y$-axis is the
    corresponding Z-score for the enrichment test (see
    Sec.~\ref{sec:grouptest}), shown green if the Z-score is positive,
    in red if it is negative. Non-transparent squares are for
    significant Z-scores, the grey area delimits
    $|$Z-scores$|>$1.96. High Z-score peaks in this plot represent
    enrichment for the functional category in a specific exponent
    range.  The cyan vertical line indicates the category exponent.}
\label{fig:all_bin}
\end{figure}

\clearpage
\newpage
%%% FIGURA S10 - exponent category with largest z-score (after the
%%% enrichment test) vs category exponent
\begin{figure}[!htbp]
\centering
  \includegraphics[width=0.65\textwidth]{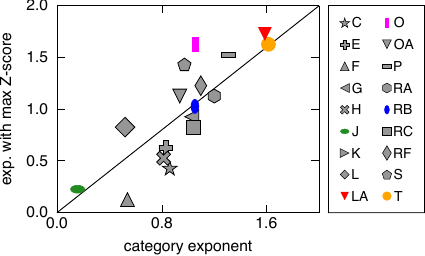}
  \caption{ \textbf{Comparison of the category exponent with the
      exponent corresponding to the maximum Z-score in the enrichment
      test in Fig.~\ref{fig:2Bpfam} (see Sec.~\ref{sec:grouptest}).}
    The black line is the $y=x$ line. Correspondence with this line
    indicates clear association between the functional category and
    the scaling exponent range. The panel on the right shows the
    association between symbols and category codes (see
    Table~\ref{tab:TableS0sym} for the corresponding category name).}
\label{fig:realExp_maxZexp}
\end{figure}

%%%% FIGURA S10 - Category scaling
%\begin{figure}[!htbp]
%\centering
%  \includegraphics[width=0.6\textwidth]{FigS10_cat_scaling}
%  \caption{The scaling of family abundance with size is not the same
%  for sub-categories within the same category.  The left panel show
%  that the number of domains with a given functional annotation
%  scales as a power-law with genome size.  This scaling can be
%  superlinear (exponent larger than one, e.g., for trascription
%  factors, red points), linear (e.g., in metabolism, blue points) or
%  sublinear (exponent lower than one, e.g., for translation, green
%  points).  The domains belonging to each category can be subdivided
%  in smaller groups (sub-categories). The right panels show that
%  different metabolic sub-categories also scale as a power-law with
%  genome size. Interestingly, the value of the exponent is not equal
%  to the one of the category they belong to and is different between
%  different sub-categories. For instance, the number of domains
%  involved in nucleotide transport and metabolism scale sublinearly
%  with the genome size (exponent 0.5), while metabolic domains scale
%  linearly with size (i.e., the exponent is equal to $1$).  }
%\label{fig:catscaling}
%\end{figure}

\clearpage
\newpage
%%%%%%%%%%% FIGURES FOR PFAM 

%%% FIGURA S11 - Scaling categories supfam vs clan
\begin{figure}[!htbp]
\centering
\includegraphics[width=0.4\textwidth]{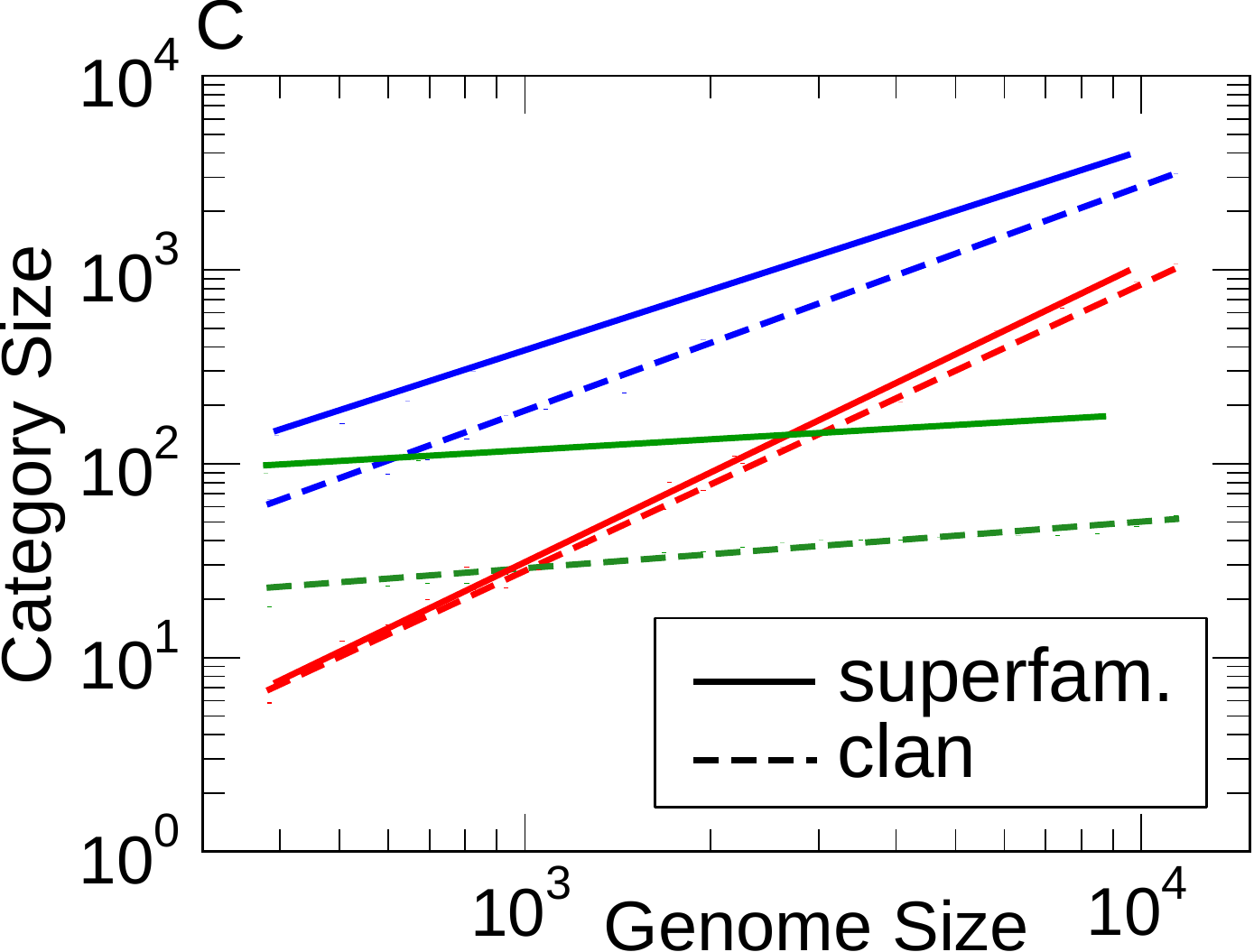}
\caption{\textbf{Comparison of the scaling of functional categories
    with gneome size in SUPERFAMILY database and Pfam database.} The
  figure shows the fitted power-law scaling of the SUPERFAMILY (solid
  lines) and Pfam (dashed lines) categories. The chosen categories are
  DNA-binding (red), Translation (green) and all metabolic categories
  (blue). The categories show similar exponents (but different
  prefactors) in the two databases.}
\label{fig:sfVSclan}
\end{figure}

%%% FIGURA S12 - Fig 2 fatta per PFAM
\begin{figure}[!htbp]
\centering
 \includegraphics[width=0.6\textwidth]{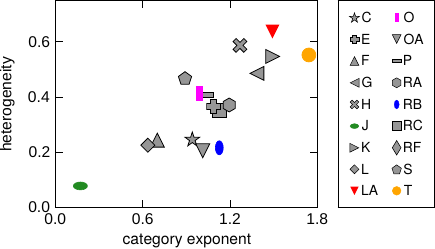}
 \caption{ \textbf{Functional categories of Pfam clans with faster
     scaling exponents contain clans with more heterogeneous scaling
     laws.} Same as Figure 2A of the Main Text, for Pfam
   clans. Heterogeneity is quantified by the mean deviation between
   the clan scaling exponents and the category exponent. The plot
   reports heterogeneity scores for different functional categories,
   plotted as a function of the category exponents. Each symbol
   corresponds to a different functional category. Only categories
   with more than 5 clans are shown. The right panel shows the
   association between symbols and category codes (see
   Table~\ref{tab:TableS0sym} for the corresponding category name).}
 \label{fig:HcVSexp_pfam}
\end{figure}

\clearpage
\newpage
%%% FIGURA S13 - exponent category with largest z-score (after the enrichment test) vs category exponent
\begin{figure}[!htbp]
\centering
  \includegraphics[width=0.6\textwidth]{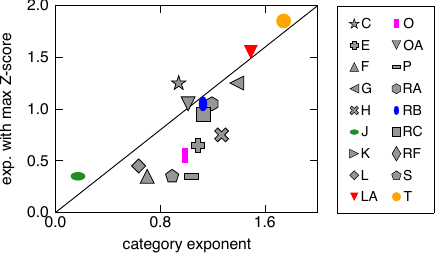}
  \caption{\textbf{Functional enrichment of sets of Pfam clans with
      similar scaling exponents.} Same as
    Fig.~\ref{fig:realExp_maxZexp} for Pfam clans. Comparison of the
    category exponent with the exponent corresponding to the maximum
    Z-score in the enrichment test (see
    Sec.~\ref{sec:grouptest}). Clans are grouped into sliding bins
    according to the value of their scaling exponent and tested for
    enrichment against each functional category. The exponent
    corresponding to the maximal value of Z-score ($y$-axis) is
    compared to the category scaling exponent ($x$-axis). The black
    line is the $y=x$ line. Correspondence with this line indicates
    clear association between the functional category and the scaling
    exponent range. The right panel shows the association between
    symbols and category codes (see Table~\ref{tab:TableS0sym} for the
    corresponding category name).}
\label{fig:realExp_maxZexp_pfam}
\end{figure}

\clearpage

\section*{Supplementary Tables}

\singlespacing

%%% TABLE S0 - Functional category name and corresponding code
\begin{longtable}[ht!!]{ c c l  }
\caption{\textbf{Symbols and codes used to identify functional categories.}} \\
\hline \hline \rule{0pt}{2.5ex}
symbol & category code & category name \\  [0.5ex]
\hline
\includegraphics[width=0.02\textwidth]{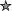} &	C	& Energy  \\ [0.5ex]
\includegraphics[width=0.02\textwidth]{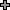} &	E	& Amino acids met./tr.  \\ [0.5ex]
\includegraphics[width=0.02\textwidth]{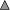} &	F	& Nucleotide met./tr.  \\ [0.5ex]
\includegraphics[width=0.02\textwidth]{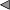} &	G	& Carbohydrate met./tr.  \\ [0.5ex]
\includegraphics[width=0.02\textwidth]{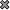} &	H	& Coenzyme met./tr.  \\ [0.5ex]
\includegraphics[width=0.02\textwidth]{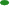} &	J	& Translation \\ [0.5ex]
\includegraphics[width=0.02\textwidth]{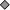} &	L	& DNA replication  \\ [0.5ex]
\includegraphics[width=0.02\textwidth]{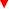} &	LA	& DNA binding  \\ [0.5ex]
\includegraphics[width=0.009\textwidth]{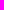} &	O	& Protein modification  \\ [0.5ex]
\includegraphics[width=0.02\textwidth]{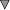} &	OA	& Proteases  \\ [0.5ex]
\includegraphics[width=0.02\textwidth]{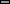} &	P	& Ion met./tr.  \\ [0.5ex]
\includegraphics[width=0.02\textwidth]{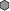} &	RA	& Redox  \\ [0.5ex]
\includegraphics[width=0.013\textwidth]{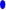} &	RB	& transferase  \\ [0.5ex]
\includegraphics[width=0.02\textwidth]{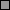} &	RC	& Other enzymes  \\ [0.5ex]
\includegraphics[width=0.013\textwidth]{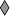} &	RF	& Transport  \\ [0.5ex]
\includegraphics[width=0.02\textwidth]{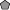} &	S	& Unknown function  \\ [0.5ex]
\includegraphics[width=0.02\textwidth]{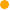} &	T	& Signal transductino  \\ [0.5ex]
\bottomrule
\label{tab:TableS0sym} \\
\end{longtable}

\clearpage
\newpage
\begin{table}[ht!!]
\centering
\caption{\textbf{Spearman correlations among family parameters.} The
  table reports Spearman correlation coefficients between sets of
  family parameters, comparing biological/evolutionary and abundance
  properties. Each row describes biological
  parameters: for the Superfamily database we used the foldability
  (quantified by size-corrected contact order, SMCO~\cite{Debes2013})
  and the diversity of EC-numbers (quantifying the functional
  plasticity of a given family, see Section~\ref{sec:ECnum})
  associated with families. For Pfam families, we considered the
  Hidden Markov Model sequence length (Hmm length) and the
  evolutionary rate (retrieved from~\cite{Ndhlovu2015}). The
  parameters listed in columns are the exponent and prefactor of the
  family scaling law ($\beta_i$ and $\alpha_i$ respectively), the mean
  family abundance calculated over all genomes ($\langle a_i \rangle$)
  and the ratio between the average relative abundance (see definition
  of frequency in Section~\ref{sec:nullmodel}) and family occurrence
  ($f_i / o_i$). Relevant correlations are found for the diversity in
  EC numbers restricted to metabolic families and the scaling exponent
  $\beta_i$, as well as with the mean and relative family
  abundance. Family abundance and scaling exponent are also correlated
(Spearman 0.72). } 
\begin{tabular}{ c|l|c|c|c|c}
\hline
\hline
\rule{0pt}{2.5ex}
\emph{database} & \emph{parameters} & $\beta_i$ & $\alpha_i$	& $\langle a_i \rangle$ & $f_i / o_i$ \\ [0.5ex]
\hline
\multirow{5}{*}{SUPFAM} & SMCO & $-0.06$ & $0.07$ & $0.04$ & $0.04$ \\ [0.5ex]
\cmidrule{2-6}
& EC numbers & \multirow{2}{*}{$0.22$} & \multirow{2}{*}{$-0.14$} & \multirow{2}{*}{$0.40$} & \multirow{2}{*}{$0.35$} \\ [0.5ex]
& (not met. families) & & & & \\ [0.5ex]
\cmidrule{2-6}
& EC numbers & \multirow{2}{*}{$0.64$} & \multirow{2}{*}{$-0.50$} &  \multirow{2}{*}{$0.77$} & \multirow{2}{*}{$0.74$} \\ [0.5ex]
& (met. families) & & & & \\ [0.5ex]
\hline
\multirow{2}{*}{PFAM} & Hmm length & $0.13$ & $-0.13$ & $0.10$ & $0.10$ \\ [0.5ex]
\cmidrule{2-6}
& Ka/Ks & $0.11$ & $-0.12$ & $0.03$ & $0.05$ \\ [0.5ex]
\bottomrule
\end{tabular}
\label{tab:TableCorrelations}
\end{table}

%% "A table contaning all the families and categories exponents is available in the SI"
\clearpage
\newpage
\begin{longtable}[ht!!]{c|l|c|c}
\caption{\textbf{Scaling exponent of functional categories.} The table
  reports the scaling exponent $\beta_c$ of all functional categories
  examined, both for superfamilies (SUPFAM column) and clans (PFAM
  column). The error associated with the exponent is calculated as the
  root mean square deviation of the logarithm of the category
  abundance across all genomes from the estimated scaling law (see
  Methods).} \\ 
\hline \hline \rule{0pt}{2.5ex}
cat. code & category name & $\beta_c \pm  \sigma_{\beta_{c}} $~(Supfam) &  $\beta_c \pm  \sigma_{\beta_{c}} $~(Pfam) \\  [0.5ex]
\hline
A	&	RNA binding, met./tr.		& $0.4	\pm	0.1 $	& $	0.21 \pm 0.10$ \\ [0.5ex]
B	&	Chromatin structure			& $0.6	\pm	0.3 $	& $--$ \\ [0.5ex]
C	&	Energy						& $0.8	\pm	0.1 $	& $	0.94 \pm 0.09$\\ [0.5ex]
CA	&	E-transfer					& $1.8	\pm	0.3 $	& $	1.73 \pm 0.34$\\ [0.5ex]
CB	&	Photosynthesis				& $0.9	\pm	0.4 $	& $	0.56 \pm 0.28$\\ [0.5ex]
D	&	Cell cycle, Apoptosis		& $1.2	\pm	0.2 $	& $--$\\ [0.5ex]
E	&	Amino acids m/tr			& $0.9	\pm	0.2 $	& $	1.09 \pm 0.17$\\ [0.5ex]
EA	&	Nitrogen m/tr				& $1.8	\pm	0.2 $	& $--$\\ [0.5ex]
F	&	Nucleotide m/tr				& $0.5	\pm	0.1 $	& $	0.70 \pm 0.17$\\ [0.5ex]
G	&	Carbohydrate m/tr			& $1.0	\pm	0.2 $	& $	1.38 \pm 0.31$\\ [0.5ex]
GA	&	Polysaccharide m/tr			& $1.1	\pm	0.2 $	& $	1.13 \pm 0.20$\\ [0.5ex]
H	&	Coenzyme m/tr				& $0.9	\pm	0.1 $	& $	1.27 \pm 0.18$\\ [0.5ex]
HA	&	Small molecule binding		& $0.9	\pm	0.1 $	& $	0.74 \pm 0.06$\\ [0.5ex]
HD	&	Receptor activity			& $1.3	\pm	0.5 $	& $	0.25 \pm 0.20$\\ [0.5ex]
HE	&	Ligand binding				& $0.3	\pm	0.1 $	& $	1.19 \pm 0.31$\\ [0.5ex]
I	&	Lipid m/tr					& $1.2	\pm	0.2 $	& $	1.31 \pm 0.20$\\ [0.5ex]
IA	&	Phospholipid m/tr			& $0.6	\pm	0.3 $	& $	0.77 \pm 0.25$\\ [0.5ex]
J	&	Translation					& $0.16	\pm	0.03$	& $	0.17 \pm 0.05$ \\ [0.5ex]
K	&	Transcription				& $0.9	\pm 0.1 $	& $	0.98 \pm 0.19$\\ [0.5ex]
L	&	DNA replication/repair		& $0.5	\pm 0.1 $	& $	0.63 \pm 0.07$\\ [0.5ex]
LA	&	DNA-binding					& $1.5	\pm 0.1 $	& $	1.49 \pm 0.13$\\ [0.5ex]
LB	&	RNA processing				& $0.2	\pm 0.1 $	& $	0.07 \pm 0.12$\\ [0.5ex]
M	&	Cell envelope m/tr			& $0.7	\pm 0.4 $	& $--$\\ [0.5ex]
MA 	&	Cell adhesion				& $1.3	\pm 0.3 $	& $	1.49 \pm 0.29$\\ [0.5ex]
N	&	Cell motility				& $0.7	\pm 0.3 $	& $	0.48 \pm 0.20$\\ [0.5ex]
O	&	Protein modification		& $1.1	\pm 0.1 $	& $	0.99 \pm 0.13$\\ [0.5ex]
OA	&	Proteases					& $1.0	\pm 0.1 $	& $	1.01 \pm 0.10$\\ [0.5ex]
OB	&	Kinases/phosphatases		& $1.3	\pm 0.2 $	& $	1.02 \pm 0.31$\\ [0.5ex]
P	&	Ion m/tr					& $1.3	\pm 0.1 $	& $	1.04 \pm 0.17$\\ [0.5ex]
Q	&	Secondary metabolism		& $1.5	\pm 0.2 $	& $	2.01 \pm 0.29$\\ [0.5ex]
R	&	General						& $1.2	\pm 0.2 $	& $	1.61 \pm 0.23$  \\ [0.5ex]
RA	&	Redox						& $1.2	\pm 0.1 $	& $	1.19 \pm 0.13$\\ [0.5ex]
RB	&	Transferases				& $1.1	\pm 0.1 $	& $	1.13 \pm 0.09$\\ [0.5ex]
RC	&	Other enzymes				& $1.1	\pm 0.1 $	& $	1.13 \pm 0.06$\\ [0.5ex]
RD	&	Protein interaction			& $1.0	\pm 0.2 $	& $	0.93 \pm 0.19$\\ [0.5ex]
RF	&	Transport					& $1.1	\pm 0.2 $	& $	1.10 \pm 0.16$\\ [0.5ex]
S	&	Unknown function			& $1.0	\pm 0.1 $	& $	0.89 \pm 0.13$\\ [0.5ex]
SB	&	Toxins/defense				& $1.1	\pm 0.3 $	& $	1.28 \pm 0.29$\\ [0.5ex]
T	&	Signal transduction			& $1.6	\pm 0.2 $	& $	1.74 \pm 0.20$\\ [0.5ex]
TA	&	Other regulatory function	& $1.1	\pm 0.2 $	& $	0.55 \pm 0.24$\\ [0.5ex]
\bottomrule
\label{tab:TableS1cat} \\
\end{longtable}
\clearpage
\newpage

\begin{longtable}[ht!!]{c|l|c}
\caption{\textbf{Scaling exponent of superfamilies from the SUPERFAMILY
  database.} The abundance of a (super)family scales as a power law of
  the genome size with family-dependent scaling exponents
  $\beta_i$. Each row corresponds to a domain family and shows its
  scaling exponent along with its error (see Methods) and the category
  to which the family belongs (category code). Families corresponding
  to the same functional category are ordered in decreasing order of
  abundance. }
\\
\hline \hline \rule{0pt}{2.5ex}
cat. code & family name & $\beta_i \pm  \sigma_{\beta_{i}} $ \\  [0.5ex]
\hline
A	 & Alpha-L RNA-binding motif														& $ 0.2 	\pm 	0.1 $ \\ [0.5ex]
A	 & PUA domain-like																	& $ 0.7	\pm 	0.2 $ \\ [0.5ex]
C	 & 6-phosphogluconate dehydrogenase C-terminal domain-like							& $ 1.2	\pm 	0.2 $ \\ [0.5ex]
\multirow{ 2}{*}{C}	 & Glyceraldehyde-3-phosphate dehydrogenase-like, 				& \multirow{ 2}{*}{$ 1.0	\pm 	0.2 $} \\ [0.5ex]
& C-terminal domain	 & \\ [0.5ex]
C	 & Phosphoenolpyruvate/pyruvate domain												& $ 1.0	\pm 	0.2 $ \\ [0.5ex]
C	 & SIS domain																		& $ 0.7	\pm 	0.2 $ \\ [0.5ex]
C	 & LeuD/IlvD-like																	& $ 0.9	\pm 	0.2 $ \\ [0.5ex]
C	 & Enolase C-terminal domain-like													& $ 0.8	\pm 	0.2 $ \\ [0.5ex]
C	 & Transmembrane di-heme cytochromes												& $ 0.5	\pm 	0.3 $ \\ [0.5ex]
C	 & Aconitase iron-sulfur domain														& $ 0.5	\pm 	0.2 $ \\ [0.5ex]
C	 & Cytochrome c oxidase subunit I-like												& $ 0.5	\pm 	0.2 $ \\ [0.5ex]
\multirow{ 2}{*}{C}	 & UDP-glucose/						& \multirow{ 2}{*}{$ 0.5	\pm 	0.2 $} \\ [0.5ex]
& GDP-mannose dehydrogenase C-terminal domain	 & \\ [0.5ex]
C	 & Citrate synthase																	& $ 0.6	\pm 	0.2 $ \\ [0.5ex]
C	 & PEP carboxykinase-like															& $ 0.2	\pm 	0.2 $ \\ [0.5ex]
C	 & Cytochrome c oxidase subunit III-like											& $ 0.4	\pm 	0.2 $ \\ [0.5ex]
C	 & PK C-terminal domain-like														& $ 0.2	\pm 	0.1 $ \\ [0.5ex]
\multirow{ 2}{*}{C}	 & Enzyme I of the PEP:sugar phosphotransferase  					& \multirow{ 2}{*}{$ 0.4	\pm 	0.2 $} \\
& system HPr-binding (sub)domain	& \\[0.5ex]
C	 & Cytochrome c oxidase subunit II-like, transmembrane region						& $ 0.3	\pm 	0.2 $ \\ [0.5ex]
CA	 & Cytochrome c																		& $ 1.0	\pm 	0.5 $ \\ [0.5ex]
CA	 & Acyl-CoA dehydrogenase C-terminal domain-like									& $ 2.0	\pm 	0.4 $ \\ [0.5ex]
CA	 & FMN-dependent nitroreductase-like												& $ 0.8	\pm 	0.3 $ \\ [0.5ex]
CA	 & ISP domain																		& $ 1.2	\pm 	0.3 $ \\ [0.5ex]
CA	 & Sulfite reductase hemoprotein (SiRHP), domains 2 and 4							& $ 0.9	\pm 	0.2 $ \\ [0.5ex]
\multirow{ 2}{*}{CA}	 & Succinate dehydrogenase/fumarate reductase flavoprotein, 	& \multirow{ 2}{*}{$ 0.7	\pm 	0.2 $} \\
& catalytic domain & \\[0.5ex]
CB	 & PRC-barrel domain																& $ 0.7	\pm 	0.3 $ \\ [0.5ex]
D	 & Rhodanese/Cell cycle control phosphatase											& $ 1.1	\pm 	0.3 $ \\ [0.5ex]
E	 & ACT-like																			& $ 0.8	\pm 	0.2 $ \\ [0.5ex]
\multirow{ 2}{*}{E}	 & Tryptophan synthase beta subunit-like						& \multirow{ 2}{*}{$ 1.0	\pm 	0.2 $} \\ [0.5ex]
&  PLP-dependent enzymes  & \\ [0.5ex]
E	 & Carbamate kinase-like															& $ 0.5	\pm 	0.1 $ \\ [0.5ex]
E	 & PLP-binding barrel																& $ 0.7	\pm 	0.1 $ \\ [0.5ex]
E	 & Glutamine synthetase/guanido kinase												& $ 0.7	\pm 	0.2 $ \\ [0.5ex]
E	 & L-aspartase-like																	& $ 0.7	\pm 	0.2 $ \\ [0.5ex]
E	 & Diaminopimelate epimerase-like													& $ 0.5	\pm 	0.2 $ \\ [0.5ex]
E	 & Alanine racemase C-terminal domain-like											& $ 0.5	\pm 	0.2 $ \\ [0.5ex]
E	 & Aspartate/glutamate racemase														& $ 0.3	\pm 	0.2 $ \\ [0.5ex]
E	 & Arginase/deacetylase																& $ 0.8	\pm 	0.2 $ \\ [0.5ex]
E	 & Aspartate/ornithine carbamoyltransferase											& $ 0.3	\pm 	0.1 $ \\ [0.5ex]
E	 & Serine metabolism enzymes domain													& $ 0.3	\pm 	0.2 $ \\ [0.5ex]
E	 & Chorismate mutase II																& $ 0.4	\pm 	0.2 $ \\ [0.5ex]
EA	 & RmlC-like cupins																	& $ 1.8	\pm 	0.2 $ \\ [0.5ex]
F	 & Ribonuclease H-like																& $ 0.7	\pm 	0.3 $ \\ [0.5ex]
F	 & Adenine nucleotide alpha hydrolases-like											& $ 0.9	\pm 	0.1 $ \\ [0.5ex]
F	 & Nucleotidylyl transferase														& $ 0.17	\pm 0.05    $ \\ [0.5ex]
F	 & PRTase-like																		& $ 0.4	\pm 	0.1 $ \\ [0.5ex]
F	 & Nucleotidyltransferase															& $ 0.7	\pm 	0.2 $ \\ [0.5ex]
F	 & Pseudouridine synthase															& $ 0.3	\pm 	0.1 $ \\ [0.5ex]
F	 & Ribulose-phoshate binding barrel													& $ 0.5	\pm 	0.2 $ \\ [0.5ex]
F	 & Tetrahydrobiopterin biosynthesis enzymes-like									& $ 0.4	\pm 	0.2 $ \\ [0.5ex]
F	 & Purine and uridine phosphorylases												& $ 0.3	\pm 	0.2 $ \\ [0.5ex]
F	 & Nucleotidyltransferase substrate binding subunit/domain							& $ 0.2	\pm 	0.2 $ \\ [0.5ex]
F	 & Nicotinate/Quinolinate PRTase C-terminal domain-like								& $ 0.4	\pm 	0.2 $ \\ [0.5ex]
\multirow{ 2}{*}{F}	 & Nucleoside phosphorylase/				& \multirow{ 2}{*}{$ 0.3	\pm 	0.2 $} \\
& phosphoribosyltransferase catalytic domain & \\[0.5ex]
\multirow{ 2}{*}{F}	 & Nucleoside phosphorylase/				& \multirow{ 2}{*}{$ 0.4	\pm 	0.2 $} \\
& phosphoribosyltransferase N-terminal domain & \\[0.5ex]
F	 & NadA-like																		& $ 0.00	\pm 0.03    $ \\ [0.5ex]
G	 & (Trans)glycosidases																& $ 1.4	\pm 	0.4 $ \\ [0.5ex]
G	 & Aldolase																			& $ 0.9	\pm 	0.2 $ \\ [0.5ex]
G	 & Phosphoglucomutase, first 3 domains												& $ 0.4	\pm 	0.1 $ \\ [0.5ex]
G	 & Galactose-binding domain-like													& $ 0.8	\pm 	0.5 $ \\ [0.5ex]
G	 & Six-hairpin glycosidases															& $ 1.2	\pm 	0.4 $ \\ [0.5ex]
G	 & Duplicated hybrid motif															& $ 0.5	\pm 	0.2 $ \\ [0.5ex]
G	 & Xylose isomerase-like															& $ 1.0	\pm 	0.3 $ \\ [0.5ex]
G	 & Carbohydrate phosphatase															& $ 0.6	\pm 	0.2 $ \\ [0.5ex]
G	 & HIT-like																			& $ 0.6	\pm 	0.2 $ \\ [0.5ex]
G	 & Phosphoglucomutase, C-terminal domain											& $ 0.4	\pm 	0.1 $ \\ [0.5ex]
G	 & PK beta-barrel domain-like														& $ 0.9	\pm 	0.2 $ \\ [0.5ex]
G	 & HPr-like																			& $ 0.2	\pm 	0.2 $ \\ [0.5ex]
GA	 & UDP-Glycosyltransferase/glycogen phosphorylase									& $ 1.0	\pm 	0.2 $ \\ [0.5ex]
GA	 & Pectin lyase-like																& $ 1.3	\pm 	0.4 $ \\ [0.5ex]
GA	 & Glycosyl hydrolase domain														& $ 0.8	\pm 	0.3 $ \\ [0.5ex]
GA	 & Barwin-like endoglucanases														& $ 0.5	\pm 	0.2 $ \\ [0.5ex]
H	 & Glutathione synthetase ATP-binding domain-like									& $ 0.9	\pm 	0.1 $ \\ [0.5ex]
H	 & Acyl-CoA dehydrogenase NM domain-like											& $ 2.0	\pm 	0.4 $ \\ [0.5ex]
H	 & PreATP-grasp domain																& $ 0.7	\pm 	0.1 $ \\ [0.5ex]
H	 & Single hybrid motif																& $ 0.8	\pm 	0.2 $ \\ [0.5ex]
H	 & FMN-binding split barrel															& $ 1.2	\pm 	0.2 $ \\ [0.5ex]
H	 & Riboflavin synthase domain-like													& $ 1.0	\pm 	0.2 $ \\ [0.5ex]
H	 & Succinyl-CoA synthetase domains													& $ 0.5	\pm 	0.2 $ \\ [0.5ex]
H	 & YrdC/RibB																		& $ 0.5	\pm 	0.2 $ \\ [0.5ex]
H	 & Molybdenum cofactor biosynthesis proteins										& $ 0.6	\pm 	0.2 $ \\ [0.5ex]
H	 & Dihydrofolate reductase-like														& $ 0.6	\pm 	0.2 $ \\ [0.5ex]
H	 & UROD/MetE-like																	& $ 0.5	\pm 	0.3 $ \\ [0.5ex]
H	 & Dihydropteroate synthetase-like													& $ 0.4	\pm 	0.2 $ \\ [0.5ex]
H	 & Cobalamin (vitamin B12)-binding domain											& $ 0.4	\pm 	0.3 $ \\ [0.5ex]
H	 & Activating enzymes of the ubiquitin-like proteins								& $ 0.4	\pm 	0.2 $ \\ [0.5ex]
H	 & Nicotinate/Quinolinate PRTase N-terminal domain-like								& $ 0.4	\pm 	0.2 $ \\ [0.5ex]
H	 & Glutamine synthetase, N-terminal domain											& $ 0.6	\pm 	0.2 $ \\ [0.5ex]
H	 & Peptide deformylase																& $ 0.2	\pm 	0.2 $ \\ [0.5ex]
H	 & RibA-like																		& $ 0.4	\pm 	0.2 $ \\ [0.5ex]
H	 & MoeA C-terminal domain-like														& $ 0.2	\pm 	0.2 $ \\ [0.5ex]
H	 & ApbE-like																		& $ 0.3	\pm 	0.2 $ \\ [0.5ex]
HA	 & P-loop containing nucleoside triphosphate hydrolases								& $ 0.71	\pm 0.08    $ \\ [0.5ex]
HA	 & NAD(P)-binding Rossmann-fold domains												& $ 1.4	\pm 	0.1 $ \\ [0.5ex]
HA	 & FAD/NAD(P)-binding domain														& $ 1.3	\pm 	0.2 $ \\ [0.5ex]
HA	 & Thiamin diphosphate-binding fold (THDP-binding)									& $ 0.9	\pm 	0.1 $ \\ [0.5ex]
HA	 & FAD-binding domain																& $ 1.0	\pm 	0.2 $ \\ [0.5ex]
HA	 & Nucleotide-binding domain														& $ 0.8	\pm 	0.2 $ \\ [0.5ex]
HA	 & Sensory domain-like																& $ 0.7	\pm 	0.4 $ \\ [0.5ex]
\multirow{ 2}{*}{HD}	 & Methyl-accepting chemotaxis protein (MCP) 						& \multirow{ 2}{*}{$ 1.0	\pm 	0.5 $} \\ [0.5ex]
& signaling domain  & \\ [0.5ex]
HD	 & PhoU-like																		& $ 0.3	\pm 	0.2 $ \\ [0.5ex]
HE	 & TGS-like																			& $ 0.3	\pm 	0.1 $ \\ [0.5ex]
I	 & Thioesterase/thiol ester dehydrase-isomerase										& $ 1.5	\pm 	0.2 $ \\ [0.5ex]
\multirow{ 2}{*}{I}	 & Probable ACP-binding domain of 						& \multirow{ 2}{*}{$ 1.0	\pm 	0.3 $} \\ [0.5ex]
& malonyl-CoA ACP transacylase & \\ [0.5ex]
I	 & Creatinase/prolidase N-terminal domain											& $ 0.5	\pm 	0.2 $ \\ [0.5ex]
\multirow{ 2}{*}{I}	 & Prokaryotic lipoproteins and 					& \multirow{ 2}{*}{$ 0.4	\pm 	0.2 $} \\ [0.5ex]
& lipoprotein localization factors & \\ [0.5ex]
IA	 & PLC-like phosphodiesterases														& $ 0.6	\pm 	0.2 $ \\ [0.5ex]
J	 & Ribosomal protein S5 domain 2-like												& $ 0.30	\pm 0.07    $ \\ [0.5ex]
J	 & Translation proteins																& $ 0.26	\pm 0.06    $ \\ [0.5ex]
J	 & EF-G C-terminal domain-like														& $ 0.30	\pm 0.09    $ \\ [0.5ex]
J	 & Sm-like ribonucleoproteins														& $ 0.8	\pm 	0.3 $ \\ [0.5ex]
J	 & Triger factor/SurA peptide-binding domain-like									& $ 0.3	\pm 	0.2 $ \\ [0.5ex]
J	 & ValRS/IleRS/LeuRS editing domain													& $ 0.03	\pm 0.04    $ \\ [0.5ex]
J	 & Release factor																	& $ 0.2	\pm 	0.1 $ \\ [0.5ex]
J	 & L30e-like																		& $ 0.3	\pm 	0.2 $ \\ [0.5ex]
J	 & EF-Tu/eEF-1alpha/eIF2-gamma C-terminal domain									& $ 0.5	\pm 	0.2 $ \\ [0.5ex]
J	 & S13-like H2TH domain																& $ 0.3	\pm 	0.2 $ \\ [0.5ex]
J	 & NusB-like																		& $ 0.3	\pm 	0.1 $ \\ [0.5ex]
J	 & ClpS-like																		& $ 0.4	\pm 	0.1 $ \\ [0.5ex]
J	 & Ribosome binding protein Y (YfiA homologue)										& $ 0.3	\pm 	0.1 $ \\ [0.5ex]
K	 & Tetracyclin repressor-like, C-terminal domain									& $ 2.4	\pm 	0.3 $ \\ [0.5ex]
K	 & LexA/Signal peptidase															& $ 0.6	\pm 	0.2 $ \\ [0.5ex]
K	 & Poly A polymerase C-terminal region-like											& $ 0.2	\pm 	0.2 $ \\ [0.5ex]
K	 & GreA transcript cleavage protein, N-terminal domain								& $ 0.2	\pm 	0.1 $ \\ [0.5ex]
K	 & CYTH-like phosphatases															& $ 0.3	\pm 	0.1 $ \\ [0.5ex]
L	 & Nucleic acid-binding proteins													& $ 0.31	\pm 0.07    $ \\ [0.5ex]
L	 & DNA breaking-rejoining enzymes													& $ 1.0	\pm 	0.3 $ \\ [0.5ex]
L	 & Nudix																			& $ 1.2	\pm 	0.2 $ \\ [0.5ex]
L	 & RuvA domain 2-like																& $ 0.4	\pm 	0.1 $ \\ [0.5ex]
L	 & Restriction endonuclease-like													& $ 0.8	\pm 	0.3 $ \\ [0.5ex]
L	 & DNA/RNA polymerases																& $ 0.8	\pm 	0.2 $ \\ [0.5ex]
L	 & DNA-glycosylase																	& $ 0.7	\pm 	0.2 $ \\ [0.5ex]
L	 & DNase I-like																	& $ 0.8	\pm 	0.2 $ \\ [0.5ex]
\multirow{ 2}{*}{L}	 & DNA polymerase III clamp loader subunits, 						& \multirow{ 2}{*}{$ 0.2	\pm 	0.1 $} \\ [0.5ex]
& C-terminal domain  & \\[0.5ex]
L	 & Resolvase-like																	& $ 0.4	\pm 	0.4 $ \\ [0.5ex]
L	 & Uracil-DNA glycosylase-like														& $ 0.6	\pm 	0.2 $ \\ [0.5ex]
L	 & GIY-YIG endonuclease															& $ 0.4	\pm 	0.2 $ \\ [0.5ex]
L	 & DNA ligase/mRNA capping enzyme, catalytic domain								& $ 0.7	\pm 	0.2 $ \\ [0.5ex]
L	 & HRDC-like																		& $ 0.4	\pm 	0.2 $ \\ [0.5ex]
L	 & N-terminal domain of MutM-like DNA repair proteins								& $ 0.4	\pm 	0.2 $ \\ [0.5ex]
L	 & TRCF domain-like																& $ 0.02	\pm 0.03    $ \\ [0.5ex]
LA	 & Winged helix DNA-binding domain													& $ 1.8	\pm 	0.2 $ \\ [0.5ex]
LA	 & Homeodomain-like																& $ 2.2	\pm 	0.3 $ \\ [0.5ex]
LA	 & lambda repressor-like DNA-binding domains										& $ 1.4	\pm 	0.3 $ \\ [0.5ex]
\multirow{ 2}{*}{LA}	 & C-terminal effector domain of 					& \multirow{ 2}{*}{$ 1.8	\pm 	0.2 $} \\ [0.5ex]
& the bipartite response regulators  & \\[0.5ex]
LA	 & Periplasmic binding protein-like I												& $ 1.4	\pm 	0.4 $ \\ [0.5ex]
LA	 & Putative DNA-binding domain														& $ 1.1	\pm 	0.2 $ \\ [0.5ex]
\multirow{ 2}{*}{LA}	 & Fatty acid responsive transcription factor FadR,				& \multirow{ 2}{*}{$ 1.9	\pm 	0.3 $} \\
& C-terminal domain & \\[0.5ex]
LA	 & Glucocorticoid receptor-like (DNA-binding domain)								& $ 0.4	\pm 	0.2 $ \\ [0.5ex]
LA	 & TrpR-like																		& $ 0.4	\pm 	0.3 $ \\ [0.5ex]
LA	 & Ribbon-helix-helix																& $ 0.8	\pm 	0.3 $ \\ [0.5ex]
LA	 & IHF-like DNA-binding proteins													& $ 0.3	\pm 	0.3 $ \\ [0.5ex]
LA	 & ParB/Sulfiredoxin																& $ 0.6	\pm 	0.3 $ \\ [0.5ex]
LA	 & KorB DNA-binding domain-like													& $ 0.4	\pm 	0.3 $ \\ [0.5ex]
LB	 & EPT/RTPC-like																	& $ 0.3	\pm 	0.1 $ \\ [0.5ex]
M	 & OmpA-like																		& $ 0.9	\pm 	0.4 $ \\ [0.5ex]
MA	 & vWA-like																		& $ 1.2	\pm 	0.3 $ \\ [0.5ex]
MA	 & Pili subunits																	& $ 0.8	\pm 	0.4 $ \\ [0.5ex]
MA	 & PGBD-like																		& $ 0.7	\pm 	0.3 $ \\ [0.5ex]
MA	 & Hedgehog/DD-peptidase															& $ 0.5	\pm 	0.2 $ \\ [0.5ex]
\multirow{ 2}{*}{O}	 & ATPase domain of HSP90 chaperone/		& \multirow{ 2}{*}{$ 1.5	\pm 	0.2 $} \\
& DNA topoisomerase II/ histidine kinase	 & \\[0.5ex]
O	 & GroES-like																		& $ 1.6	\pm 	0.3 $ \\ [0.5ex]
O	 & FKBP-like																		& $ 0.6	\pm 	0.2 $ \\ [0.5ex]
O	 & Chaperone J-domain																& $ 0.5	\pm 	0.2 $ \\ [0.5ex]
O	 & Cyclophilin-like																& $ 0.9	\pm 	0.2 $ \\ [0.5ex]
O	 & Double Clp-N motif																& $ 0.6	\pm 	0.2 $ \\ [0.5ex]
O	 & HSP20-like chaperones															& $ 0.6	\pm 	0.2 $ \\ [0.5ex]
O	 & GroEL equatorial domain-like													& $ 0.2	\pm 	0.2 $ \\ [0.5ex]
O	 & GroEL apical domain-like														& $ 0.2	\pm 	0.2 $ \\ [0.5ex]
O	 & Peptide methionine sulfoxide reductase											& $ 0.2	\pm 	0.2 $ \\ [0.5ex]
O	 & GroEL-intermediate domain like													& $ 0.2	\pm 	0.1 $ \\ [0.5ex]
OA	 & ClpP/crotonase																	& $ 1.0	\pm 	0.2 $ \\ [0.5ex]
OA	 & Zn-dependent exopeptidases														& $ 1.0	\pm 	0.2 $ \\ [0.5ex]
OA	 & Metallo-dependent phosphatases													& $ 1.0	\pm 	0.2 $ \\ [0.5ex]
OA	 & Metalloproteases ("zincins"), catalytic domain									& $ 0.8	\pm 	0.3 $ \\ [0.5ex]
OA	 & LuxS/MPP-like metallohydrolase													& $ 0.5	\pm 	0.3 $ \\ [0.5ex]
OA	 & Cysteine proteinases															& $ 1.0	\pm 	0.3 $ \\ [0.5ex]
OA	 & Bacterial exopeptidase dimerisation domain										& $ 1.1	\pm 	0.3 $ \\ [0.5ex]
OA	 & Trypsin-like serine proteases													& $ 1.0	\pm 	0.3 $ \\ [0.5ex]
OA	 & Creatinase/aminopeptidase														& $ 0.5	\pm 	0.1 $ \\ [0.5ex]
OA	 & HSP40/DnaJ peptide-binding domain												& $ 0.3	\pm 	0.2 $ \\ [0.5ex]
OA	 & DPP6 N-terminal domain-like														& $ 0.6	\pm 	0.4 $ \\ [0.5ex]
OA	 & Subtilisin-like																	& $ 0.8	\pm 	0.3 $ \\ [0.5ex]
OA	 & Rhomboid-like																	& $ 0.5	\pm 	0.2 $ \\ [0.5ex]
OA	 & Macro domain-like																& $ 0.3	\pm 	0.2 $ \\ [0.5ex]
OA	 & Tricorn protease N-terminal domain												& $ 0.4	\pm 	0.2 $ \\ [0.5ex]
OB	 & Protein kinase-like (PK-like)													& $ 1.2	\pm 	0.3 $ \\ [0.5ex]
OB	 & PP2C-like																		& $ 0.8	\pm 	0.3 $ \\ [0.5ex]
OB	 & Phosphohistidine domain															& $ 0.7	\pm 	0.2 $ \\ [0.5ex]
OB	 & Phosphotyrosine protein phosphatases I											& $ 0.6	\pm 	0.2 $ \\ [0.5ex]
OB	 & Acylphosphatase/BLUF domain-like												& $ 0.3	\pm 	0.2 $ \\ [0.5ex]
P	 & Periplasmic binding protein-like II												& $ 1.6	\pm 	0.3 $ \\ [0.5ex]
P	 & MFS general substrate transporter												& $ 1.4	\pm 	0.3 $ \\ [0.5ex]
P	 & Multidrug resistance efflux transporter EmrE									& $ 1.3	\pm 	0.3 $ \\ [0.5ex]
P	 & HlyD-like secretion proteins													& $ 1.5	\pm 	0.4 $ \\ [0.5ex]
P	 & Ferritin-like																	& $ 1.0	\pm 	0.2 $ \\ [0.5ex]
P	 & Cupredoxins																		& $ 1.1	\pm 	0.3 $ \\ [0.5ex]
P	 & Calcium ATPase, transduction domain A											& $ 0.6	\pm 	0.2 $ \\ [0.5ex]
P	 & Calcium ATPase, transmembrane domain M											& $ 0.6	\pm 	0.2 $ \\ [0.5ex]
P	 & TrkA C-terminal domain-like														& $ 0.6	\pm 	0.3 $ \\ [0.5ex]
P	 & HMA, heavy metal-associated domain												& $ 0.4	\pm 	0.2 $ \\ [0.5ex]
P	 & Band 7/SPFH domain																& $ 0.5	\pm 	0.2 $ \\ [0.5ex]
P	 & Fe-S cluster assembly (FSCA) domain-like										& $ 0.4	\pm 	0.2 $ \\ [0.5ex]
P	 & Voltage-gated potassium channels												& $ 0.6	\pm 	0.2 $ \\ [0.5ex]
P	 & Magnesium transport protein CorA, transmembrane region							& $ 0.5	\pm 	0.2 $ \\ [0.5ex]
P	 & CorA soluble domain-like														& $ 0.5	\pm 	0.2 $ \\ [0.5ex]
P	 & Clc chloride channel															& $ 0.3	\pm 	0.2 $ \\ [0.5ex]
Q	 & Dimeric alpha+beta barrel														& $ 2.1	\pm 	0.3 $ \\ [0.5ex]
Q	 & Clavaminate synthase-like														& $ 1.6	\pm 	0.3 $ \\ [0.5ex]
Q	 & Concanavalin A-like lectins/glucanases											& $ 0.8	\pm 	0.4 $ \\ [0.5ex]
Q	 & Terpenoid synthases																& $ 0.7	\pm 	0.2 $ \\ [0.5ex]
\multirow{ 2}{*}{Q}	 & Homo-oligomeric flavin-containing 						& \multirow{ 2}{*}{$ 0.4	\pm 	0.2 $} \\ [0.5ex]
& Cys decarboxylases, HFCD  & \\ [0.5ex]
R	 & Bet v1-like																		& $ 1.5	\pm 	0.4 $ \\ [0.5ex]
R	 & Helical backbone metal receptor													& $ 0.9	\pm 	0.3 $ \\ [0.5ex]
R	 & ADC-like																		& $ 1.0	\pm 	0.3 $ \\ [0.5ex]
R	 & ARM repeat																		& $ 0.6	\pm 	0.3 $ \\ [0.5ex]
\multirow{ 2}{*}{R}	 & Peripheral subunit-binding domain of 2-oxo 		& \multirow{ 2}{*}{$ 0.3	\pm 	0.2 $} \\
& acid dehydrogenase complex	 & \\[0.5ex]
R	 & Pentein																			& $ 0.4	\pm 	0.2 $ \\ [0.5ex]
R	 & JAB1/MPN domain																	& $ 0.2	\pm 	0.2 $ \\ [0.5ex]
RA	 & Thioredoxin-like																& $ 1.1	\pm 	0.2 $ \\ [0.5ex]
RA	 & 4Fe-4S ferredoxins																& $ 0.8	\pm 	0.4 $ \\ [0.5ex]
RA	 & Metallo-hydrolase/oxidoreductase												& $ 1.1	\pm 	0.2 $ \\ [0.5ex]
\multirow{ 2}{*}{RA}	 & Glyoxalase/Bleomycin resistance protein/		& \multirow{ 2}{*}{$ 2.1	\pm 	0.3 $} \\
& Dihydroxybiphenyl dioxygenase	 & \\[0.5ex]
RA	 & ALDH-like																		& $ 1.5	\pm 	0.2 $ \\ [0.5ex]
RA	 & 2Fe-2S ferredoxin-like															& $ 1.0	\pm 	0.3 $ \\ [0.5ex]
RA	 & Flavoproteins																	& $ 0.9	\pm 	0.3 $ \\ [0.5ex]
RA	 & alpha-helical ferredoxin														& $ 0.9	\pm 	0.3 $ \\ [0.5ex]
RA	 & FAD-linked reductases, C-terminal domain										& $ 1.4	\pm 	0.3 $ \\ [0.5ex]
RA	 & Formate/glycerate dehydrogenase catalytic domain-like							& $ 1.0	\pm 	0.2 $ \\ [0.5ex]
RA	 & NAD(P)-linked oxidoreductase													& $ 1.3	\pm 	0.3 $ \\ [0.5ex]
RA	 & Isocitrate/Isopropylmalate dehydrogenase-like									& $ 0.6	\pm 	0.2 $ \\ [0.5ex]
RA	 & Aminoacid dehydrogenase-like, N-terminal domain									& $ 0.7	\pm 	0.1 $ \\ [0.5ex]
\multirow{ 2}{*}{RA}	 & FAD/NAD-linked reductases, 					& \multirow{ 2}{*}{$ 0.7	\pm 	0.2 $} \\ [0.5ex]
& dimerisation (C-terminal) domain	  & \\ [0.5ex]
RA	 & Formate dehydrogenase/DMSO reductase, domains 1-3								& $ 1.0	\pm 	0.3 $ \\ [0.5ex]
RA	 & Ferredoxin reductase-like, C-terminal NADP-linked domain						& $ 1.0	\pm 	0.3 $ \\ [0.5ex]
RA	 & Dehydroquinate synthase-like													& $ 0.7	\pm 	0.3 $ \\ [0.5ex]
RA	 & Inosine monophosphate dehydrogenase (IMPDH)										& $ 0.5	\pm 	0.2 $ \\ [0.5ex]
RA	 & Acid phosphatase/Vanadium-dependent haloperoxidase								& $ 0.7	\pm 	0.2 $ \\ [0.5ex]
RA	 & FAD-linked oxidases, C-terminal domain											& $ 1.0	\pm 	0.3 $ \\ [0.5ex]
RA	 & Sulfite reductase, domains 1 and 3												& $ 0.6	\pm 	0.2 $ \\ [0.5ex]
\multirow{ 2}{*}{RA}	 & Succinate dehydrogenase/		& \multirow{ 2}{*}{$ 0.5	\pm 	0.2 $} \\ [0.5ex]
& fumarate reductase flavoprotein C-terminal domain & \\[0.5ex]
RA	 & LDH C-terminal domain-like														& $ 0.3	\pm 	0.2 $ \\ [0.5ex]
RA	 & FAD-linked oxidoreductase														& $ 0.6	\pm 	0.2 $ \\ [0.5ex]
RB	 & S-adenosyl-L-methionine-dependent methyltransferases							& $ 0.9	\pm 	0.1 $ \\ [0.5ex]
RB	 & PLP-dependent transferases														& $ 1.2	\pm 	0.1 $ \\ [0.5ex]
RB	 & Acyl-CoA N-acyltransferases (Nat)												& $ 1.7	\pm 	0.2 $ \\ [0.5ex]
RB	 & Nucleotide-diphospho-sugar transferases											& $ 0.9	\pm 	0.2 $ \\ [0.5ex]
RB	 & Class I glutamine amidotransferase-like											& $ 1.0	\pm 	0.1 $ \\ [0.5ex]
RB	 & CoA-dependent acyltransferases													& $ 1.0	\pm 	0.5 $ \\ [0.5ex]
RB	 & NagB/RpiA/CoA transferase-like													& $ 1.1	\pm 	0.2 $ \\ [0.5ex]
RB	 & TK C-terminal domain-like														& $ 0.7	\pm 	0.2 $ \\ [0.5ex]
RB	 & FabD/lysophospholipase-like														& $ 1.1	\pm 	0.3 $ \\ [0.5ex]
RB	 & Tetrapyrrole methylase															& $ 1.0	\pm 	0.3 $ \\ [0.5ex]
RB	 & Glycerol-3-phosphate (1)-acyltransferase										& $ 0.7	\pm 	0.2 $ \\ [0.5ex]
RB	 & Formyltransferase																& $ 0.6	\pm 	0.1 $ \\ [0.5ex]
RB	 & D-aminoacid aminotransferase-like PLP-dependent enzymes							& $ 0.5	\pm 	0.2 $ \\ [0.5ex]
RB	 & 4'-phosphopantetheinyl transferase												& $ 0.5	\pm 	0.2 $ \\ [0.5ex]
\multirow{ 2}{*}{RB}	 & Methylated DNA-protein cysteine methyltransferase, 			& \multirow{ 2}{*}{$ 0.8	\pm 	0.2 $} \\
& C-terminal domain & \\[0.5ex]
RB	 & Methylated DNA-protein cysteine methyltransferase domain						& $ 0.6	\pm 	0.2 $ \\ [0.5ex]
RB	 & Homocysteine S-methyltransferase												& $ 0.2	\pm 	0.2 $ \\ [0.5ex]
RC	 & alpha/beta-Hydrolases															& $ 1.6	\pm 	0.3 $ \\ [0.5ex]
RC	 & Actin-like ATPase domain														& $ 0.7	\pm 	0.1 $ \\ [0.5ex]
RC	 & HAD-like																		& $ 0.9	\pm 	0.2 $ \\ [0.5ex]
RC	 & Thiolase-like																	& $ 1.3	\pm 	0.3 $ \\ [0.5ex]
RC	 & Radical SAM enzymes																& $ 0.7	\pm 	0.3 $ \\ [0.5ex]
RC	 & Acetyl-CoA synthetase-like														& $ 1.9	\pm 	0.3 $ \\ [0.5ex]
RC	 & Metallo-dependent hydrolases													& $ 1.3	\pm 	0.2 $ \\ [0.5ex]
RC	 & HD-domain/PDEase-like															& $ 0.7	\pm 	0.3 $ \\ [0.5ex]
RC	 & beta-lactamase/transpeptidase-like												& $ 0.9	\pm 	0.2 $ \\ [0.5ex]
RC	 & Trimeric LpxA-like enzymes														& $ 0.7	\pm 	0.2 $ \\ [0.5ex]
RC	 & Lysozyme-like																	& $ 1.0	\pm 	0.2 $ \\ [0.5ex]
RC	 & Composite domain of metallo-dependent hydrolases								& $ 1.4	\pm 	0.2 $ \\ [0.5ex]
RC	 & N-terminal nucleophile aminohydrolases (Ntn hydrolases)							& $ 1.1	\pm 	0.2 $ \\ [0.5ex]
RC	 & Ribokinase-like																	& $ 0.9	\pm 	0.2 $ \\ [0.5ex]
RC	 & Alkaline phosphatase-like														& $ 1.0	\pm 	0.3 $ \\ [0.5ex]
RC	 & DHS-like NAD/FAD-binding domain													& $ 1.2	\pm 	0.2 $ \\ [0.5ex]
RC	 & Phospholipase D/nuclease														& $ 0.8	\pm 	0.2 $ \\ [0.5ex]
RC	 & Glycoside hydrolase/deacetylase													& $ 1.1	\pm 	0.2 $ \\ [0.5ex]
RC	 & Cytidine deaminase-like															& $ 0.6	\pm 	0.1 $ \\ [0.5ex]
RC	 & LysM domain																		& $ 0.3	\pm 	0.4 $ \\ [0.5ex]
RC	 & SGNH hydrolase																	& $ 0.9	\pm 	0.3 $ \\ [0.5ex]
RC	 & PurM N-terminal domain-like														& $ 0.2	\pm 	0.1 $ \\ [0.5ex]
RC	 & PurM C-terminal domain-like														& $ 0.2	\pm 	0.1 $ \\ [0.5ex]
RC	 & Phosphoglycerate mutase-like													& $ 0.9	\pm 	0.3 $ \\ [0.5ex]
RC	 & Galactose mutarotase-like														& $ 0.6	\pm 	0.3 $ \\ [0.5ex]
RC	 & Carbon-nitrogen hydrolase														& $ 0.8	\pm 	0.2 $ \\ [0.5ex]
RC	 & PHP domain-like																	& $ 0.6	\pm 	0.2 $ \\ [0.5ex]
RC	 & Enolase N-terminal domain-like													& $ 0.9	\pm 	0.3 $ \\ [0.5ex]
RC	 & Quinoprotein alcohol dehydrogenase-like											& $ 0.9	\pm 	0.3 $ \\ [0.5ex]
RC	 & all-alpha NTP pyrophosphatases													& $ 0.5	\pm 	0.2 $ \\ [0.5ex]
RC	 & FAH																				& $ 1.3	\pm 	0.3 $ \\ [0.5ex]
RC	 & PFL-like glycyl radical enzymes													& $ 0.3	\pm 	0.3 $ \\ [0.5ex]
RC	 & Amidase signature (AS) enzymes													& $ 0.8	\pm 	0.3 $ \\ [0.5ex]
RC	 & Isochorismatase-like hydrolases													& $ 1.2	\pm 	0.3 $ \\ [0.5ex]
RC	 & L,D-transpeptidase catalytic domain-like										& $ 0.9	\pm 	0.3 $ \\ [0.5ex]
RC	 & Chorismate lyase-like															& $ 0.8	\pm 	0.3 $ \\ [0.5ex]
RC	 & MoCo carrier protein-like														& $ 0.5	\pm 	0.2 $ \\ [0.5ex]
RC	 & NAD kinase																		& $ 0.5	\pm 	0.2 $ \\ [0.5ex]
RC	 & ADC synthase																	& $ 0.4	\pm 	0.2 $ \\ [0.5ex]
RC	 & Folate-binding domain															& $ 0.6	\pm 	0.2 $ \\ [0.5ex]
RC	 & AraD-like aldolase/epimerase													& $ 0.7	\pm 	0.2 $ \\ [0.5ex]
RC	 & FMT C-terminal domain-like														& $ 0.3	\pm 	0.2 $ \\ [0.5ex]
RC	 & IlvD/EDD N-terminal domain-like													& $ 0.7	\pm 	0.2 $ \\ [0.5ex]
RC	 & Chelatase																		& $ 0.4	\pm 	0.2 $ \\ [0.5ex]
RC	 & Aminomethyltransferase beta-barrel domain										& $ 0.6	\pm 	0.2 $ \\ [0.5ex]
\multirow{ 2}{*}{RC}	 & 2-isopropylmalate synthase LeuA, 				& \multirow{ 2}{*}{$1 0.3	\pm 	0.2 $} \\ [0.5ex]
& allosteric (dimerisation) domain & \\ [0.5ex]
RC	 & CNF1/YfiH-like putative cysteine hydrolases										& $ 0.2	\pm 	0.2 $ \\ [0.5ex]
RC	 & Nqo1 middle domain-like															& $ 0.2	\pm 	0.2 $ \\ [0.5ex]
RC	 & beta-carbonic anhydrase, cab													& $ 0.7	\pm 	0.2 $ \\ [0.5ex]
RC	 & N-acetylmuramoyl-L-alanine amidase-like											& $ 0.3	\pm 	0.2 $ \\ [0.5ex]
RC	 & post-HMGL domain-like															& $ 0.4	\pm 	0.2 $ \\ [0.5ex]
RC	 & Nqo1C-terminal domain-like														& $ 0.2	\pm 	0.2 $ \\ [0.5ex]
RC	 & DmpA/ArgJ-like																	& $ 0.4	\pm 	0.2 $ \\ [0.5ex]
RC	 & Riboflavin kinase-like															& $ 0.0	\pm 	0.0 $ \\ [0.5ex]
RC	 & LigT-like																		& $ 0.3	\pm 	0.2 $ \\ [0.5ex]
RD	 & TPR-like																		& $ 1.2	\pm 	0.4 $ \\ [0.5ex]
RD	 & FMN-linked oxidoreductases														& $ 1.0	\pm 	0.1 $ \\ [0.5ex]
RD	 & Nqo1 FMN-binding domain-like													& $ 0.3	\pm 	0.2 $ \\ [0.5ex]
RF	 & Multidrug efflux transporter AcrB transmembrane domain							& $ 1.2	\pm 	0.3 $ \\ [0.5ex]
\multirow{ 2}{*}{RF}	 & Multidrug efflux transporter AcrB pore domain;	& \multirow{ 2}{*}{$ 1.2	\pm 	0.4 $} \\
& PN1, PN2, PC1 and PC2 subdomains & \\[0.5ex]
\multirow{ 2}{*}{RF}	 & Multidrug efflux transporter AcrB TolC docking domain;	& \multirow{ 2}{*}{$ 1.2	\pm 	0.4 $} \\
& DN and DC subdomains	& \\ [0.5ex]
RF	 & CBS-domain																		& $ 0.9	\pm 	0.2 $ \\ [0.5ex]
RF	 & ABC transporter transmembrane region											& $ 0.7	\pm 	0.3 $ \\ [0.5ex]
RF	 & NTF2-like																		& $ 1.6	\pm 	0.3 $ \\ [0.5ex]
RF	 & Outer membrane efflux proteins (OEP)											& $ 1.2	\pm 	0.3 $ \\ [0.5ex]
RF	 & ABC transporter involved in vitamin B12 uptake, BtuC							& $ 0.8	\pm 	0.3 $ \\ [0.5ex]
RF	 & Rudiment single hybrid motif													& $ 0.8	\pm 	0.2 $ \\ [0.5ex]
\multirow{ 2}{*}{RF}	 & Mechanosensitive channel protein MscS (YggB), 					& \multirow{ 2}{*}{$ 0.6	\pm 	0.3 $} \\ [0.5ex]
& C-terminal domain & \\ [0.5ex]
\multirow{ 2}{*}{RF}	 & Mechanosensitive channel protein MscS (YggB),			& \multirow{ 2}{*}{$ 0.6	\pm 	0.2 $} \\
& transmembrane region	 & \\[0.5ex]
RF	 & Proton glutamate symport protein												& $ 0.3	\pm 	0.3 $ \\ [0.5ex]
RF	 & Ammonium transporter															& $ 0.3	\pm 	0.2 $ \\ [0.5ex]
S	 & Sigma2 domain of RNA polymerase sigma factors									& $ 1.4	\pm 	0.3 $ \\ [0.5ex]
S	 & ACP-like																		& $ 1.3	\pm 	0.4 $ \\ [0.5ex]
S	 & alpha/beta knot																	& $ 0.3	\pm 	0.1 $ \\ [0.5ex]
S	 & E set domains																	& $ 1.0	\pm 	0.3 $ \\ [0.5ex]
S	 & MOP-like																		& $ 1.0	\pm 	0.3 $ \\ [0.5ex]
S	 & PIN domain-like																	& $ 0.7	\pm 	0.3 $ \\ [0.5ex]
S	 & Anti-sigma factor antagonist SpoIIaa											& $ 1.1	\pm 	0.3 $ \\ [0.5ex]
S	 & YjgF-like																		& $ 1.3	\pm 	0.2 $ \\ [0.5ex]
S	 & HCP-like																		& $ 0.4	\pm 	0.4 $ \\ [0.5ex]
S	 & ITPase-like																		& $ 0.4	\pm 	0.1 $ \\ [0.5ex]
S	 & MoaD/ThiS																		& $ 0.5	\pm 	0.2 $ \\ [0.5ex]
S	 & YbaK/ProRS associated domain													& $ 0.6	\pm 	0.2 $ \\ [0.5ex]
S	 & Sporulation related repeat														& $ 0.3	\pm 	0.2 $ \\ [0.5ex]
S	 & GatB/YqeY motif																	& $ 0.3	\pm 	0.1 $ \\ [0.5ex]
SB	 & AhpD-like																		& $ 1.5	\pm 	0.3 $ \\ [0.5ex]
T	 & CheY-like																		& $ 1.7	\pm 	0.2 $ \\ [0.5ex]
T	 & PYP-like sensor domain (PAS domain)												& $ 2.0	\pm 	0.5 $ \\ [0.5ex]
T	 & Homodimeric domain of signal transducing histidine kinase						& $ 1.6	\pm 	0.3 $ \\ [0.5ex]
T	 & Nucleotide cyclase																& $ 1.6	\pm 	0.4 $ \\ [0.5ex]
T	 & GAF domain-like																	& $ 1.7	\pm 	0.3 $ \\ [0.5ex]
T	 & PDZ domain-like																	& $ 0.5	\pm 	0.2 $ \\ [0.5ex]
T	 & EAL domain-like																	& $ 0.9	\pm 	0.4 $ \\ [0.5ex]
T	 & cAMP-binding domain-like														& $ 1.4	\pm 	0.3 $ \\ [0.5ex]
T	 & Histidine-containing phosphotransfer domain, HPT domain							& $ 1.2	\pm 	0.3 $ \\ [0.5ex]
T	 & GlnB-like																		& $ 0.6	\pm 	0.2 $ \\ [0.5ex]
T	 & Mss4-like																		& $ 0.6	\pm 	0.3 $ \\ [0.5ex]
\multirow{ 2}{*}{TA}	 & Sigma3 and sigma4 domains of 						& \multirow{ 2}{*}{$ 1.1	\pm 	0.2 $} \\ [0.5ex]
& RNA polymerase sigma factors  & \\ [0.5ex]
TA	 & OsmC-like																		& $ 1.0	\pm 	0.2 $ \\ [0.5ex]
TA	 & CinA-like																		& $ 0.3	\pm 	0.1 $ \\ [0.5ex]
\bottomrule
\label{tab:TableS1fam} \\
\end{longtable}
\clearpage
\newpage

\begin{longtable}[ht!!]{c|l|c}
\caption{\textbf{Scaling exponent of Pfam clans.} The abundance of a clan
  scales as a power law of the genome size with family-dependent
  scaling exponents $\beta_i$. Each row of the table corresponds to a
  clan and shows its scaling exponent along with its error (see
  Methods) and the corresponding functional category (category
  code). Clans associated to the same functional category are ordered
  in decreasing order of abundance.  } \\
\hline \hline \rule{0pt}{2.5ex}
cat. code & clan name & $\beta_i \pm  \sigma_{\beta_{i}} $ \\  [0.5ex]
\hline
A	&	S4 domain superfamily													& $0.29 \pm	0.15$ \\ [0.5ex]
C	&	Pyruvate kinase-like TIM barrel superfamily								& $1.16 \pm	0.19$ \\ [0.5ex]
C	&	6-phosphogluconate dehydrogenase C-terminal-like superfamily			& $1.10 \pm	0.16$ \\ [0.5ex]
C	&	SIS domain fold															& $0.76 \pm	0.21$ \\ [0.5ex]
C	&	Transmembrane di-heme cytochrome superfamily							& $0.82 \pm	0.24$ \\ [0.5ex]
C	&	Enolase like TIM barrel													& $1.13 \pm	0.28$ \\ [0.5ex]
C	&	PFK-like superfamily													& $0.50 \pm	0.23$ \\ [0.5ex]
C	&	LeuD/IlvD-like															& $0.52 \pm	0.16$ \\ [0.5ex]
CA	&	Cytochrome c superfamily												& $0.88 \pm	0.45$ \\ [0.5ex]
CA	&	Acyl-CoA dehydrogenase, C-terminal domain-like							& $1.95 \pm	0.42$ \\ [0.5ex]
CA	&	Rieske-like iron-sulphur domain											& $1.15 \pm	0.30$ \\ [0.5ex]
CA	&	FMN-dependent nitroreductase-like										& $0.78 \pm	0.24$ \\ [0.5ex]
CB	&	PRC-barrel like superfamily												& $0.57 \pm	0.28$ \\ [0.5ex]
E	&	ACT-like domain															& $0.70 \pm	0.20$ \\ [0.5ex]
E	&	gamma-glutamylcysteine synthetase/glutamine synthetase clan				& $0.91 \pm	0.23$ \\ [0.5ex]
E	&	DAP epimerase superfamily												& $0.51 \pm	0.17$ \\ [0.5ex]
E	&	Arginase/deacetylase superfamily										& $0.74 \pm	0.24$ \\ [0.5ex]
E	&	Aspartate/glutamate racemase superfamily								& $0.74 \pm	0.23$ \\ [0.5ex]
F	&	Ribonuclease H-like superfamily											& $0.78 \pm	0.29$ \\ [0.5ex]
F	&	Nucleotidyltransferase superfamily										& $0.73 \pm	0.19$ \\ [0.5ex]
F	&	PRPP synthetase-associated protein 1									& $0.43 \pm	0.15$ \\ [0.5ex]
F	&	Tetrahydrobiopterin biosynthesis-like enzyme superfamily				& $0.43 \pm	0.20$ \\ [0.5ex]
F	&	Nucleotidyltransferase substrate binding domain							& $0.37 \pm	0.26$ \\ [0.5ex]
F	&	Purine and uridine phosphorylase superfamily							& $0.44 \pm	0.21$ \\ [0.5ex]
F	&	dUTPase like superfamily												& $0.22 \pm	0.15$ \\ [0.5ex]
G	&	Tim barrel glycosyl hydrolase superfamily								& $1.32 \pm	0.42$ \\ [0.5ex]
G	&	Six-hairpin glycosidase superfamily										& $1.13 \pm	0.36$ \\ [0.5ex]
G	&	Galactose-binding domain-like superfamily								& $0.84 \pm	0.46$ \\ [0.5ex]
G	&	inositol polyphosphate 1 phosphatase like superfamily					& $0.65 \pm	0.24$ \\ [0.5ex]
G	&	HIT superfamily															& $0.55 \pm	0.23$ \\ [0.5ex]
GA	&	Glycosyl transferase clan GT-B											& $0.98 \pm	0.19$ \\ [0.5ex]
GA	&	Pectate lyase-like beta helix											& $1.49 \pm	0.52$ \\ [0.5ex]
GA	&	Glycosyl hydrolase domain superfamily									& $0.87 \pm	0.26$ \\ [0.5ex]
GA	&	Double Psi beta barrel glucanase										& $0.55 \pm	0.19$ \\ [0.5ex]
H	&	ATP-grasp superfamily													& $0.88 \pm	0.15$ \\ [0.5ex]
H	&	Acyl-coenzyme A oxidase/dehydrogenase N-terminal						& $1.86 \pm	0.42$ \\ [0.5ex]
H	&	Riboflavin synthase/Ferredoxin reductase FAD binding domain				& $0.90 \pm	0.23$ \\ [0.5ex]
H	&	FMN-binding split barrel superfamily									& $1.52 \pm	0.27$ \\ [0.5ex]
H	&	Dihydrofolate reductase-like											& $0.65 \pm	0.23$ \\ [0.5ex]
H	&	Release factor superfamily												& $0.29 \pm	0.09$ \\ [0.5ex]
H	&	Succinyl-CoA synthetase flavodoxin domain superfamily					& $0.34 \pm	0.18$ \\ [0.5ex]
\multirow{ 2}{*}{HA}	&	P-loop containing nucleoside 			& \multirow{ 2}{*}{$0.70 \pm	0.07$} \\ [0.5ex]
& triphosphate hydrolase superfamily  & \\[0.5ex]
HA	&	PCMH-like FAD binding													& $1.37 \pm	0.25$ \\ [0.5ex]
HD	&	PhoU-like superfamily													& $0.41 \pm	0.20$ \\ [0.5ex]
HE	&	Ubiquitin superfamily													& $1.24 \pm	0.31$ \\ [0.5ex]
I	&	HotDog superfamily														& $1.60 \pm	0.24$ \\ [0.5ex]
I	&	Creatinase/prolidase N-terminal domain superfamily						& $0.54 \pm	0.20$ \\ [0.5ex]
IA	&	PLC-like phosphodiesterases												& $0.53 \pm	0.23$ \\ [0.5ex]
J	&	Ribosomal protein S5 domain 2-like superfamily							& $0.32 \pm	0.08$ \\ [0.5ex]
J	&	Transcription elongation factor G C-terminal							& $0.28 \pm	0.11$ \\ [0.5ex]
J	&	Helix-two-turns-helix superfamily										& $0.20 \pm	0.17$ \\ [0.5ex]
J	&	DALR superfamily														& $0.20 \pm	0.12$ \\ [0.5ex]
K	&	Peptidase clan SF														& $0.71 \pm	0.20$ \\ [0.5ex]
L	&	OB fold																	& $0.45 \pm	0.07$ \\ [0.5ex]
L	&	PD-(D/E)XK nuclease superfamily											& $0.58 \pm	0.23$ \\ [0.5ex]
L	&	NUDIX superfamily														& $1.19 \pm	0.21$ \\ [0.5ex]
L	&	DNA breaking-rejoining enzyme superfamily								& $0.97 \pm	0.25$ \\ [0.5ex]
L	&	His-Me finger endonuclease superfamily									& $0.78 \pm	0.28$ \\ [0.5ex]
L	&	DNase I-like															& $0.75 \pm	0.24$ \\ [0.5ex]
L	&	GIY-YIG endonuclease superfamily										& $0.36 \pm	0.22$ \\ [0.5ex]
L	&	DNA/RNA ligase superfamily												& $0.55 \pm	0.20$ \\ [0.5ex]
L	&	HRDC-like superfamily													& $0.36 \pm	0.16$ \\ [0.5ex]
LA	&	Helix-turn-helix clan													& $1.64 \pm	0.14$ \\ [0.5ex]
LA	&	Periplasmic binding protein like										& $1.49 \pm	0.38$ \\ [0.5ex]
\multirow{ 2}{*}{LA}	&	Fatty acid responsive transcription factor FadR,	& \multirow{ 2}{*}{$1.98 \pm	0.34$} \\ [0.5ex]
&  C-terminal domain	  & \\[0.5ex]
LA	&	lambda integrase N-terminal domain										& $0.49 \pm	0.22$ \\ [0.5ex]
LA	&	MetJ/Arc repressor superfamily											& $0.56 \pm	0.34$ \\ [0.5ex]
LA	&	ParB-like superfamily													& $0.70 \pm	0.26$ \\ [0.5ex]
LA	&	IHF-like DNA-binding protein supewrfamily								& $0.39 \pm	0.28$ \\ [0.5ex]
LB	&	EPT/RTPC-like superfamily												& $0.33 \pm	0.13$ \\ [0.5ex]
MA	&	Ig-like fold superfamily (E-set)										& $1.49 \pm	0.43$ \\ [0.5ex]
MA	&	von Willebrand factor type A											& $1.17 \pm	0.30$ \\ [0.5ex]
MA	&	Pilus subunit															& $0.83 \pm	0.41$ \\ [0.5ex]
MA	&	PGBD superfamily														& $0.75 \pm	0.33$ \\ [0.5ex]
MA	&	Peptidase MD															& $0.48 \pm	0.24$ \\ [0.5ex]
N	&	Flagellar motor switch family											& $0.34 \pm	0.30$ \\ [0.5ex]
O	&	GroES-like superfamily													& $1.56 \pm	0.28$ \\ [0.5ex]
O	&	FKBP-like superfamily													& $0.62 \pm	0.28$ \\ [0.5ex]
O	&	Chaperone J-domain superfamily											& $0.45 \pm	0.24$ \\ [0.5ex]
O	&	Cyclophilin-like superfamily											& $0.74 \pm	0.19$ \\ [0.5ex]
O	&	HSP20-like chaperone superfamily										& $0.65 \pm	0.24$ \\ [0.5ex]
OA	&	Peptidase clan MA														& $0.97 \pm	0.15$ \\ [0.5ex]
OA	&	ClpP/Crotonase superfamily												& $1.06 \pm	0.21$ \\ [0.5ex]
OA	&	Peptidase clan MH/MC/MF													& $1.03 \pm	0.19$ \\ [0.5ex]
OA	&	Calcineurin-like phosphoesterase superfamily							& $1.00 \pm	0.18$ \\ [0.5ex]
OA	&	Peptidase clan CA														& $1.09 \pm	0.24$ \\ [0.5ex]
OA	&	LuxS/MPP-like metallohydrolase											& $0.48 \pm	0.26$ \\ [0.5ex]
OA	&	Peptidase clan PA														& $0.79 \pm	0.26$ \\ [0.5ex]
OA	&	MACRO domain superfamily												& $0.33 \pm	0.18$ \\ [0.5ex]
OB	&	PP2C-like superfamily													& $0.97 \pm	0.34$ \\ [0.5ex]
P	&	Ferritin-like Superfamily												& $1.05 \pm	0.20$ \\ [0.5ex]
P	&	Multicopper oxidase-like domain											& $0.93 \pm	0.30$ \\ [0.5ex]
P	&	SPFH superfamily														& $0.34 \pm	0.23$ \\ [0.5ex]
P	&	SufE/NifU superfamily													& $0.23 \pm	0.16$ \\ [0.5ex]
Q	&	Dimeric alpha/beta barrel superfamily									& $2.01 \pm	0.29$ \\ [0.5ex]
R	&	Bet V 1 like															& $1.50 \pm	0.37$ \\ [0.5ex]
R	&	Acetyl-decarboxylase like superfamily									& $1.06 \pm	0.29$ \\ [0.5ex]
R	&	Helical backbone metal receptor superfamily								& $0.81 \pm	0.34$ \\ [0.5ex]
R	&	GME superfamily															& $0.44 \pm	0.21$ \\ [0.5ex]
RA	&	4Fe-4S ferredoxins														& $0.94 \pm	0.35$ \\ [0.5ex]
RA	&	Thioredoxin-like														& $1.16 \pm	0.20$ \\ [0.5ex]
RA	&	VOC superfamily															& $2.11 \pm	0.32$ \\ [0.5ex]
RA	&	Metallo-hydrolase/oxidoreductase superfamily							& $1.13 \pm	0.17$ \\ [0.5ex]
RA	&	ALDH-like superfamily													& $1.64 \pm	0.25$ \\ [0.5ex]
RA	&	2Fe-2S iron-sulfur cluster binding domain								& $1.05 \pm	0.26$ \\ [0.5ex]
RA	&	Transthyretin superfamily												& $0.94 \pm	0.53$ \\ [0.5ex]
RA	&	Flavoprotein															& $0.89 \pm	0.27$ \\ [0.5ex]
RA	&	Isocitrate/Isopropylmalate dehydrogenase-like superfamily				& $0.64 \pm	0.17$ \\ [0.5ex]
\multirow{ 2}{*}{RA}	&	Formate/glycerate dehydrogenase 		& \multirow{ 2}{*}{$0.99 \pm	0.19$} \\ [0.5ex]
&  catalytic domain-like superfamily & \\[0.5ex]
RA	&	Ferredoxin / Ferric reductase-like NAD binding							& $1.02 \pm	0.26$ \\ [0.5ex]
RA	&	Dehydroquinate synthase-like superfamily								& $0.68 \pm	0.29$ \\ [0.5ex]
RA	&	FAD-linked oxidase C-terminal domain superfamily						& $1.21 \pm	0.25$ \\ [0.5ex]
RA	&	Acid phosphatase/Vanadium-dependent haloperoxidase						& $0.77 \pm	0.24$ \\ [0.5ex]
RA	&	LDH C-terminal domain-like superfamily									& $0.23 \pm	0.25$ \\ [0.5ex]
RA	&	FAD-linked oxidoreductase												& $0.52 \pm	0.16$ \\ [0.5ex]
RB	&	PLP dependent aminotransferase superfamily								& $1.22 \pm	0.13$ \\ [0.5ex]
RB	&	N-acetyltransferase like												& $1.70 \pm	0.23$ \\ [0.5ex]
RB	&	Glycosyl transferase clan GT-A											& $0.91 \pm	0.20$ \\ [0.5ex]
RB	&	Class-I Glutamine amidotransferase superfamily							& $0.99 \pm	0.14$ \\ [0.5ex]
\multirow{2}{*}{RB}	&	Isomerase,CoA transferase \&	& \multirow{2}{*}{$1.04 \pm	0.22$} \\ [0.5ex]
& Translation initiation factor Superfamily & \\ [0.5ex]
RB	&	CoA-dependent acyltransferase superfamily								& $0.96 \pm	0.39$ \\ [0.5ex]
RB	&	Patatin/FabD/lysophospholipase-like superfamily							& $1.12 \pm	0.26$ \\ [0.5ex]
RB	&	Acyltransferase clan													& $0.67 \pm	0.28$ \\ [0.5ex]
RC	&	FAD/NAD(P)-binding Rossmann fold Superfamily							& $1.12 \pm	0.08$ \\ [0.5ex]
RC	&	Alpha/Beta hydrolase fold												& $1.54 \pm	0.29$ \\ [0.5ex]
RC	&	Actin-like ATPase Superfamily											& $0.75 \pm	0.14$ \\ [0.5ex]
RC	&	Thiolase-like Superfamily												& $1.22 \pm	0.25$ \\ [0.5ex]
RC	&	HAD superfamily															& $0.79 \pm	0.19$ \\ [0.5ex]
RC	&	Amidohydrolase superfamily												& $1.07 \pm	0.17$ \\ [0.5ex]
RC	&	Hexapeptide repeat superfamily											& $0.58 \pm	0.19$ \\ [0.5ex]
RC	&	ANL superfamily															& $1.91 \pm	0.30$ \\ [0.5ex]
RC	&	Serine beta-lactamase-like superfamily									& $0.90 \pm	0.18$ \\ [0.5ex]
RC	&	HD/PDEase superfamily													& $0.60 \pm	0.27$ \\ [0.5ex]
RC	&	NTN hydrolase superfamily												& $1.06 \pm	0.17$ \\ [0.5ex]
RC	&	Ribokinase-like superfamily												& $0.87 \pm	0.20$ \\ [0.5ex]
RC	&	Alkaline phosphatase-like												& $1.01 \pm	0.30$ \\ [0.5ex]
RC	&	Lysozyme-like superfamily												& $0.89 \pm	0.26$ \\ [0.5ex]
RC	&	LysM-like domain														& $0.53 \pm	0.34$ \\ [0.5ex]
RC	&	DHS-like NAD/FAD-binding domain											& $1.20 \pm	0.18$ \\ [0.5ex]
RC	&	Cytidine deaminase-like (CDA) superfamily								& $0.60 \pm	0.14$ \\ [0.5ex]
RC	&	Phospholipase D superfamily												& $0.55 \pm	0.28$ \\ [0.5ex]
RC	&	Histidine phosphatase superfamily										& $0.77 \pm	0.26$ \\ [0.5ex]
RC	&	Glycoside hydrolase/deacetylase superfamily								& $1.01 \pm	0.25$ \\ [0.5ex]
RC	&	Galactose Mutarotase-like superfamily									& $0.86 \pm	0.31$ \\ [0.5ex]
RC	&	SGNH hydrolase superfamily												& $0.91 \pm	0.32$ \\ [0.5ex]
RC	&	PFL-like glycyl radical enzyme superfamily								& $0.26 \pm	0.31$ \\ [0.5ex]
RC	&	Enolase N-terminal domain-like superfamily								& $1.06 \pm	0.27$ \\ [0.5ex]
\multirow{2}{*}{RC}	&	Fumarylacetoacetate hydrolase, 			& \multirow{2}{*}{$1.34 \pm	0.29$} \\ [0.5ex]
& C-terminal domain, superfamily  & \\ [0.5ex]
RC	&	L,D-transpeptidase catalytic domain										& $0.94 \pm	0.30$ \\ [0.5ex]
RC	&	Chorismate lyase/UTRA superfamily										& $0.82 \pm	0.31$ \\ [0.5ex]
RC	&	MoCo carrier protein-like superfamily									& $0.44 \pm	0.16$ \\ [0.5ex]
RC	&	Chelatase Superfamily													& $0.45 \pm	0.24$ \\ [0.5ex]
\multirow{2}{*}{RC}	&	Fumarate reductase respiratory 			& \multirow{2}{*}{$0.31 \pm	0.19$} \\ [0.5ex]
&  complex transmembrane subunits & \\ [0.5ex]
RD	&	Tetratrico peptide repeat superfamily									& $1.07 \pm	0.44$ \\ [0.5ex]
RD	&	Common phosphate binding-site TIM barrel superfamily					& $0.91 \pm	0.11$ \\ [0.5ex]
RF	&	Membrane and transport protein											& $1.01 \pm	0.30$ \\ [0.5ex]
RF	&	ABC transporter membrane domain clan									& $0.84 \pm	0.26$ \\ [0.5ex]
RF	&	NTF2-like superfamily													& $1.07 \pm	0.36$ \\ [0.5ex]
S	&	Zinc beta-ribbon														& $0.63 \pm	0.20$ \\ [0.5ex]
S	&	ACP-like superfamily													& $1.68 \pm	0.41$ \\ [0.5ex]
S	&	SPOUT Methyltransferase Superfamily										& $0.31 \pm	0.11$ \\ [0.5ex]
S	&	PIN domain superfamily													& $0.87 \pm	0.27$ \\ [0.5ex]
S	&	STAS domain superfamily													& $1.17 \pm	0.30$ \\ [0.5ex]
S	&	YjgF-like superfamily													& $1.32 \pm	0.24$ \\ [0.5ex]
\multirow{2}{*}{S}	&	Phenylalanine- and lysidine-tRNA 			& \multirow{2}{*}{$0.26 \pm	0.17$} \\ [0.5ex]
&  synthetase domain superfamily & \\ [0.5ex]
S	&	YqeY-like superfamily													& $0.27 \pm	0.14$ \\ [0.5ex]
S	&	Maf/Ham1 superfamily													& $0.30 \pm	0.17$ \\ [0.5ex]
SB	&	AhpD-like superfamily													& $1.31 \pm	0.28$ \\ [0.5ex]
ST	&	Type III antifreeze and spore coat polysaccharide						& $0.62 \pm	0.23$ \\ [0.5ex]
T	&	His Kinase A (phospho-acceptor) domain									& $1.72 \pm	0.19$ \\ [0.5ex]
T	&	CheY-like superfamily													& $1.72 \pm	0.23$ \\ [0.5ex]
T	&	PAS domain clan															& $1.92 \pm	0.44$ \\ [0.5ex]
T	&	Nucleotide cyclase superfamily											& $1.63 \pm	0.44$ \\ [0.5ex]
T	&	GAF domain-like															& $2.15 \pm	0.29$ \\ [0.5ex]
T	&	PDZ domain-like peptide-binding superfamily								& $0.59 \pm	0.18$ \\ [0.5ex]
T	&	GlnB-like superfamily													& $0.71 \pm	0.25$ \\ [0.5ex]
T	&	Src homology-3 domain													& $0.25 \pm	0.36$ \\ [0.5ex]
\bottomrule
\label{tab:TableS2clan} \\
\end{longtable}

\end{document}